\documentclass[nonatbib,12pt]{elsarticle}

\usepackage{iftex}
\ifXeTeX
  \usepackage{fontspec} % load system fonts, no fontenc/inputenc
\else
  \usepackage[T1]{fontenc}
  \usepackage[utf8]{inputenc}
\fi

\usepackage{amssymb}
\usepackage{yhmath} % widiehat and widetilde
\usepackage{multirow}
\usepackage{booktabs}
\usepackage{array}
\usepackage{tabularx}
\usepackage{threeparttable}
\usepackage{microtype}
\usepackage{graphicx}
\usepackage{longtable}
\usepackage{placeins}

\usepackage[a4paper,text={16cm,25cm}]{geometry}
\usepackage[unicode=true]{hyperref}
\hypersetup{
  colorlinks=true,
  linkcolor=blue,
  citecolor=blue,
  urlcolor=blue}

\makeatletter
\let\c@author\relax
\makeatother

\usepackage[style=apa, backend=biber, natbib=true,
giveninits=true, uniquename=false, date=year, safeinputenc]{biblatex}

\renewbibmacro{in:}{%
  \ifentrytype{article}{}{%
      \printtext{\bibstring{in}\intitlepunct}}}

\DeclareNameAlias{sortname}{family-given}

\AtEveryBibitem{%
  \clearfield{addendum}%
  \clearfield{annotation}%
  \clearfield{url}%
  \clearfield{doi}%
  \clearlist{language}%
  \clearfield{issn}%
  \clearfield{pubstate}%
  \clearfield{urlyear}%
}

\renewbibmacro*{volume+number+eid}{%
  \printfield{volume}%
  \setunit*{\addnbspace}% NEW (optional); there's also \addnbthinspace
  \printfield{number}%
  \setunit{\addcomma\space}%
  \printfield{eid}}
\DeclareFieldFormat[article]{number}{\mkbibparens{#1}}

\usepackage{xcolor}
\AtEveryCite{\color{blue}}

\usepackage{todonotes}
\usepackage{fancyhdr}
\usepackage{setspace}
\newcolumntype{Y}{>{\centering\arraybackslash}X}

\journal{Annals of Tourism Research}

\begin{document}

\begin{frontmatter}

% \title{RISE: Recovery-Informed Strategy Enhancement in Post-COVID-19 Chinese Outbound Tourism Forecasting}
\title{RECOVERY-INFORMED FORECASTING STRATEGY ENHANCEMENT}

\author[pku]{Feng Li\corref{cor}}
\ead{feng.li@gsm.pku.edu.cn}
\author[cufe]{Taozhu Ruan}
\ead{rtz12365784@outlook.com}
\cortext[cor]{Feng Li (ORCiD: 0000-0002-4248-9778) is the author for correspondence. Phone (Fax): +86 (0)10 62747602. Authors listed in alphabetical order.}

\affiliation[pku]{organization={Guanghua School of Management, Peking University},
            city={Beijing},
            postcode={100871},
            country={China}}
\affiliation[cufe]{organization={Quantitative Investment Department, Fullgoal Fund Management},
            city={Shanghai},
            postcode={200122},
            country={China}}

%% Abstract
\begin{abstract} % 120 words
We propose a three-stage framework named as Recovery-Informed Strategy Enhancement (RISE) to forecast the recovery of Chinese outbound tourism following the coronavirus disease 2019 pandemic. The framework decomposes the forecasts into three parts: the initial forecasts, the terminal forecasts and the recovery curve forecasts that connect the two points. We integrate multiple sources of information and employ forecast combination techniques in all stages, enhancing both the accuracy and robustness of recovery forecasts. Compared with conventional forecasting approaches, our framework provides a structured and transparent pipeline to integrate model-based forecasts with expert-informed judgment under structural breaks and high uncertainty. Our findings demonstrate the effectiveness of this framework, offering an adaptable tool for recovery trajectory forecasting in post-crisis contexts.
\end{abstract}

%% Graphical abstract

%% Research highlights
% \begin{highlights}
% \item The framework uses three stages to forecast tourism recovery after crises.
% \item It combines model ensembles with hierarchical reconciliation for accuracy.
% \item It leverages Baidu search data and flight info to detect early recovery signs.
% \item The method ranked first in point forecasting in a global competition.
% \end{highlights}

%% Keywords

\end{frontmatter}


% \noindent\textbf{Author Bios}
% \bigskip

% Dr. Feng Li is an Associate Professor at Guanghua School of Management, Peking University. His research focuses on Bayesian statistics, forecasting, and distributed learning, with publications in top journals and presentations at major international conferences. Mr. Taozhu Ran is an analyst at Quantitative Investment Department, Fullgoal Fund Management in Shanghai, China. His research interest includes time series forecasting and quantitative trading.

%% \newpage
\section{Introduction}
\label{sec: intro}

The tourism industry is a key driver of global economic growth, accounting for around 10\% of world GDP and providing employment for hundreds of millions of people worldwide before the COVID-19 outbreak \citep{UNTourism2019WorldTourism}. The tourism industry exhibits complex interlinkages with macroeconomic, social, and environmental systems, amplifying its role as both an engine of growth and a channel of global interdependence. International tourism fosters the flow of capital, culture, and labor across borders, thereby influencing balance-of-payments dynamics and regional development. China, as one of the largest source markets, has played an increasingly important role in international tourism, with outbound tourist departures exceeding 150 million in 2019. Tourism forecasting, which refers to the process of predicting future trends of tourist arrivals and departures, plays a vital role in effective planning and decision-making within the tourism industry, allowing stakeholders to anticipate demand fluctuations and allocate resources efficiently. Researchers have identified several forecasting methods, including traditional time series models, deep learning models, and econometric approaches, to improve prediction accuracy \citep{SongH2008TourismDemand,GohC2011MethodologicalProgress}.

The high degree of openness and dependence on mobility make the tourism industry particularly vulnerable to global shocks, such as economic crises, natural disasters, and public health emergencies. The coronavirus disease 2019 (COVID-19) severely affected the global economy, leading to widespread disruptions in various industries, with the tourism sector being one of the most affected. A significant drop in international and domestic travel forced airlines, hotels, and tour operators to close or operate at a minimal capacity \citep{SigalaM2020TourismCOVID19}. The pandemic also reshaped the structure of tourism demand, accelerating digitization, altering traveler preferences toward short-haul trips, and heightening uncertainty in travel behavior. In Mainland China's case, outbound tourism virtually halted due to stringent border control measures, flight suspensions, and quarantine requirements. As the world transitions into the post-pandemic phase, the recovery of tourism has become a crucial indicator of broader economic normalization and social resilience. Accordingly, the research objective of this paper is to develop a structured forecasting framework that systematically integrates quantitative model-based forecasts with qualitative human judgment to support tourism recovery forecasting under structural breaks and high uncertainty, enabling timely resource allocation, marketing strategies, and investment decisions for rebuilding affected sectors.

\begin{figure}
    \centering
    \includegraphics[width=1.0\textwidth]{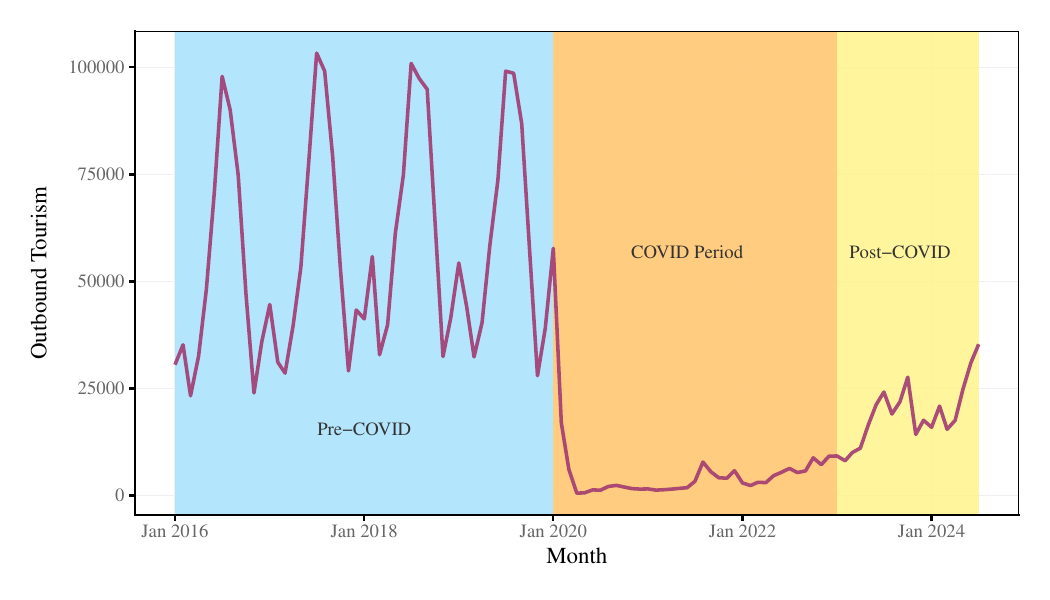}
    \caption{China's outbound tourism to Canada in three periods.}
    \label{fig: tourism_example}
\end{figure}

Nonetheless, tourism recovery forecasting in the post-pandemic era faces several intertwined challenges. The pandemic induced profound structural breaks, rendering historical data partially obsolete and undermining the assumptions of the traditional forecasting model. Figure \ref{fig: tourism_example} shows that the COVID-19 outbreak caused a structural break in the time series of tourism, making it impossible to predict the future directly from historical patterns. In such a situation, recovery trajectories are marked by high uncertainty and volatility, influenced by evolving health policies, traveler sentiment, and macroeconomic conditions. There are no universal recovery patterns across regions and market segments. Secondly, while several works acknowledge the importance of expert judgment under uncertainty, few have developed a systematic way to integrate quantitative forecasts with human judgment, even though hybrid approaches have been empirically shown to outperform purely statistical or subjective forecasts \citep{FildesR2009EffectiveForecasting,PetropoulosF2018JudgmentalSelection}.

We present a three-stage forecasting framework known as Recovery-Informed Strategy Enhancement (RISE) to forecast the recovery of Chinese outbound tourism post-COVID-19. In this framework, we divide the forecasts into three components: the initial forecasts, the terminal forecasts and the recovery curve forecasts that bridge the initial forecasts and terminal forecasts. Such decomposition allows us to apply appropriate assumptions and models across different components of the forecasts while enhancing the overall robustness by constraining the forecasts within the range of initial forecasts and terminal forecasts.

By embedding \emph{expert judgment} within a systematic quantitative framework, the RISE approach bridges the gap between statistical precision and contextual adaptability. Here, ``expert judgment'' refers to structured analyst judgment embedded through transparent scoring rules and recovery coefficients, rather than direct elicitation from external experts. This methodological innovation enhances both the robustness and interpretability of recovery predictions and offers a versatile template for forecasting under structural breaks and uncertainty.

We developed this framework for the \emph{Tourism Forecasting Competition amid COVID-19 Round II} held in 2023. In the competition, participants must predict Chinese outbound tourist arrivals for 20 selected destinations from August 2023 to July 2024, utilizing the historical data available up to February 2023. The competition featured both point-forecasting and interval-forecasting tracks. Point forecasts were evaluated based on the Mean Absolute Scaled Error. Our results ranked first in the point forecasting track and third place in the interval forecasting track of the \emph{Tourism Forecasting Competition amid COVID-19 Round II}. See, e.g., \citet{LiJ2026ForecastingTourism,ZhangH2026ForecastingChinese} for other solutions in this competition.

The contributions of our research are threefold. Theoretically, it advances the literature on post-crisis tourism forecasting by proposing a unified framework that integrates quantitative modeling with expert judgment under structural breaks and uncertainty. The RISE framework bridges the gap between statistical rigor and contextual adaptability, providing a systematic approach to incorporate human insights into model-based forecasts. Methodologically, the study develops a three-stage forecasting system that allows for dynamic integration of historical patterns, real-time signals, and heuristic adjustments, significantly enhancing forecast robustness and interpretability. Practically, this framework demonstrates superior predictive performance in forecasting China’s outbound tourism recovery after COVID-19, ranking first in the point forecasting track and third in the interval forecasting track of the Tourism Forecast Competition II. It also provides a practical and adaptable tool for policymakers and industry practitioners to evaluate tourism recovery paths and resource allocation strategies, and can be extended to other sectors facing post-crisis recovery challenges.

The remainder of this paper is structured as follows: Section \ref{sec: literature} reviews previous studies on tourism forecasting and Section \ref{sec: forecasting-framework} details the recovery informed strategy enhancement framework, outlining its three stages. Section \ref{sec: evaluation} presents the evaluation results and further conducts an ex-post analysis following the competition. Finally, Section \ref{sec: conclusion} summarizes the paper, emphasizing the methodological contributions of the research along with the policy implications of the findings.

\section{Literature review}
\label{sec: literature}

With the increasing scale and complexity of global tourism systems, tourism demand forecasting has become a critical issue for policymakers, practitioners, and researchers. In the context of COVID-19 pandemic, which has posed unprecedented challenges to tourism industry, to forecast the recovery trajectory of international tourism is extremely difficult due to the structural break and highly volatile macroeconomic conditions. Conventional model-based approaches that operate well under stable conditions perform poorly after the crisis. As a result, the literature review section will mainly focus on two parts: (i) general tourism forecasting methods, and (ii) post-crisis tourism forecasting methods.

\subsection{Tourism forecasting methods}

Under relatively stable economic and institutional conditions, among the most widely adopted approaches in tourism forecasting are time series models including the Na\"ive, autoregressive, exponential smoothing, moving average and historical average models. The Na\"ive methods are by far the most easily adopted and the most popular methods used in the tourism forecasting literature \citep{SongH2008TourismDemand}. Autoregressive Integrated Moving Average (ARIMA) models and their seasonal variations have long been regarded as a benchmark in tourism demand forecasting due to their ability to model both seasonal and non-seasonal dynamics \citep{LimC2001MonthlySeasonal,GunterU2015ForecastingInternational}. Closely related, exponential smoothing methods—including simple exponential smoothing, Holt's linear trend model, and Holt–Winters seasonal models—provide flexible and computationally efficient tools for capturing evolving trends and seasonal patterns, and have demonstrated strong forecasting performance across a wide range of tourism demand series \citep{AthanasopoulosG2008ModellingForecasting}.

Nonlinear and data-driven models have been increasingly incorporated into tourism forecasting, especially the neural network methods, with the growing volume and complexity of tourism data. Among them, the artificial neural network models are well recognized for their ability to accommodate data imperfections and to capture complex nonlinear relationships without requiring restrictive functional assumptions. These features have led to their widespread adoption in forecasting studies, particularly in contexts where demand dynamics exhibit nonlinear and unstable patterns \citep{PalmerA2006DesigningArtificial}. Meanwhile, Neural network autoregressive models extend traditional autoregressive structures by allowing for nonlinear relationships, offering improved flexibility when demand exhibits complex dynamics \citep{SalinasD2020DeepARProbabilistic}.

From the data source perspective, it is suggested that variables generated from social media and search queries are effective in forecasting tourism demand \citep{HuM2024TourismDemand,YangY2022SearchQuery,LiH2023TourismForecasting}. Stability of risk is also a significant factor that affects preferences in tourism \citep{BalazV2024StabilityRisk}.

Given the diversity of model structures and assumptions, a growing strand of literature emphasizes the value of forecast combination strategies. Rather than relying on a single best model, combining forecasts from multiple individual models has been shown to improve robustness and accuracy, particularly when demand dynamics vary over time. In tourism forecasting, forecast combination methods have consistently demonstrated superior or at least more stable performance compared to individual models, especially in cross-destination and cross-market settings \citep{AndrawisRR2011CombinationLong}.

\subsection{Post-crisis tourism forecasting methods}
In the aftermath of major crises, such as financial shocks, epidemics, or pandemics, tourism demand forecasting becomes substantially more challenging due to structural breaks, heightened uncertainty, and rapidly evolving recovery dynamics\citep{PerronP2006DealingStructural}. Typically, model-based approaches well-designed for such periods are used. Recently, the judgmental forecasting, along with expert-adjusted forecasts have also been widely used in tourism forecasting. They are integrated with quantitative models in order to provide more accurate and accountable results.

Model-based approaches to post-crisis tourism forecasting rely on well-established statistical frameworks to model structural shifts and dynamic patterns in time series data. \citet{GohC2002ModelingForecasting} combined Seasonal ARIMA model with intervention analysis for inbound passenger forecasting and examining the impact of shocks on tourism demand, including the relaxation of the issuance of outbound visitor visas, the Asian financial crisis, the handover, and the bird flu epidemic. \citet{ChuFL2008FractionallyIntegrated} incorporated fractionally integrated ARMA models into tourism forecasting after the Asian financial crisis and the September 11 event, and found that this method performed better than alternatives during crises. Seasonal ARIMA models and intervention models were also used to estimate the immediate and short-term impact of a crisis. They discovered that post-SARS tourism recovery was not a linear process but rather characterized by threshold effects and hysteresis. Different source markets exhibited significantly different recovery speeds due to variations in risk perception. Time-varying parameter model are also used to simulate the structural shift in macroeconomic atmosphere in tourism forecasting \citep{SmeralE2017TourismForecasting}. \citet{WangW2025ForecastingChinese} employed a segmented combined Vector Autoregression model with exogenous variables effectively predict Chinese outbound tourism and that accounting for the time-specific effects of exogenous factors.

\citet{SongH2010ImpactsFinancial} investigated tourism flows in Asia amid the 2008 financial and economic crisis using the autoregressive distributed lag method. By incorporating both short-run and long-run dynamics, the model allowed for a nuanced analysis of how international tourist arrivals responded to the crisis. The findings showed that the crisis significantly reduced tourism demand across Asia in the short term, but the long-run impact was less severe. \citet{LiuA2021VisitorArrivals} compared the forecasting errors of autoregressive distributed lag method, structural time-varying parameter regression, and other models in simulating tourism demand recovery after COVID-19, finding that combining multiple models substantially enhances the robustness and accuracy of forecasts.

In recent years, deep learning methods have also been applied to post-crisis tourism demand forecasting. \citet{SaaymanA2017NonlinearModels} pointed out that non-linear methods outperform other methods with a structural break in the tourism data. The literature indicates that deep learning methods more effectively capture the nonlinear pattern of a sharp drop followed by gradual recovery, supporting the development of post-pandemic multi-scenario forecasts. Other innovative approaches, such as probabilistic forecasting \citep{AthanasopoulosG2023ProbabilisticForecasts} and mixed-data frequency learning, have also been explored to capture the unprecedented uncertainty in the COVID-19 era.

Compared to model-based methods, judgmental forecasting refers to forecasting approaches that incorporate human expertise, intuition, and contextual knowledge into the forecasting process, either as a complement to or an adjustment of model-based forecasts. A substantial body of forecasting research has demonstrated that judgmental input can improve forecast accuracy, particularly when integrated with quantitative models, where statistical models provide the baseline forecasts and experts adjust them based on relevant contextual information. Numerous studies have indicated that this approach yields more accurate forecasts \citep{FildesR2009EffectiveForecasting,vanDijkD2019CombiningExpertadjusted}.

Judgmental methods and expert adjustments have played a critical role in post-crisis tourism demand forecasting, particularly in response to the COVID-19 pandemic. For example, Delphi-based judgmental adjustments made to statistical forecasts of tourism numbers improved accuracy \citep{LinVS2014AccuracyBias}. In application, \citet{LiuA2021VisitorArrivals} proposed a two-stage, three-scenario forecasting framework for inbound tourism demand across 20 countries. They developed a COVID-19 Recovery index using the Delphi method to measure pandemic impacts across destinations, which was then used to adjust baseline forecasts. \citet{KourentzesN2021VisitorArrivals} predicted international arrivals for 20 destinations via judgmental adjustment of baseline forecasts, incorporating expert assessments of policy restrictions, vaccine development, seasonality, and macroeconomic conditions. \citet{AthanasopoulosG2023ProbabilisticForecasts} further introduced a scenario-based probabilistic forecasting method grounded in large-scale expert surveys to characterize uncertainty in recovery pathways.

Nonetheless, the aforementioned literature focuses on a single aspect of the tourism forecasting challenge. Crises such as the COVID-19 pandemic are characterized by sudden onset, uncertainty, and volatility, making traditional forecasting approaches less effective in capturing recovery dynamics. There is a lack of a systematic framework that could integrate quantitative modeling, real-time external indicators, and adjustments to forecast the recovery.

\section{Forecasting framework}
\label{sec: forecasting-framework}
%% Labels are used to cross-reference an item using \ref command.
 Instead of relying on a single method or a set of methods, we propose a three-stage framework that utilizes historical outbound tourism time series alongside current external information to navigate potential pathways for tourism recovery.

 The main feature of the RISE framework is to segment the tourism recovery forecasts into three distinct parts: the initial forecasts before June 2023, the terminal forecasts in July 2024, and the recovery curve forecasts that connect the initial forecasts and the terminal forecasts from July 2023 to June 2024. The terminology definitions are listed in Table \ref{tab:recovery-def}. Unlike conventional tourism forecasting methods, RISE decomposes the forecasting task temporally and structurally, allowing different assumptions and information sets to operate at distinct stages. Importantly, our method incorporates elements of intervention analysis \citep{BoxGEP1975InterventionAnalysis} within a time series modeling framework, aimed at measuring the impact of external or structural changes (interventions) on a time series.

\begin{table}[htbp]
  \centering
  \caption{The three-stage forecast framework and term definitions.}
  \label{tab:recovery-def}
  \resizebox{\textwidth}{!}{
  \begin{tabular}{lp{12cm}p{3cm}}
    \toprule
    Name                     & Definition                                                                   & Time Period                 \\
    \midrule
    Base forecasts           & The anticipated trajectory of outbound tourism assuming no structural breaks. & January 2023 -- July 2024   \\
    Reference forecasts      & A short-term forecasting curve that can be validated in a timely manner. Consisting of forecasts after COVID-19, but at the earlier stage of the recovery. We use the period before the deadline of submitting forecasting values for the competition. & February 2023 -- June 2023  \\
    Initial forecasts    & The last forecast points for all destinations in their reference forecasting curves.                                                 & June 2023                   \\
    Terminal forecasts   & The endpoints of the forecasting horizon for all destinations.                                                  & July 2024                   \\
    Recovery coefficient &  A shrinkage factor that adjusts the last value of the base forecasts, resulting in the terminal forecast. \\
    Recovery curve & The curve that connects the initial forecast point and the terminal forecast point.     & June 2023 -- July 2024 \\
    & &\\
    \bottomrule
  \end{tabular}
  }
\end{table}

The RISE framework consists of three sequential stages designed to forecast tourism recovery under structural breaks. In the first stage, pre-2020 outbound tourism data are used to construct base forecasts that represent the counterfactual trajectory in the absence of COVID-19, from which terminal forecasts for July 2024 are extracted and adjusted by destination-specific recovery coefficients to account for lingering pandemic effects. In the second stage, initial forecasts are obtained by incorporating high-frequency external signals, including Baidu search indices and flight data, to improve short-term estimation during the early recovery period. In the final stage, recovery curve forecasts are generated by linking the initial and terminal forecasts under alternative functional forms—linear, quadratic, and logistic—to capture different recovery dynamics. Together, these stages integrate historical patterns, real-time information, and structured judgment into a coherent forecasting system, enhancing robustness and interpretability under high uncertainty. The computer code and related data to replicate these results are accessible from the author's GitHub repository at \texttt{\url{https://github.com/feng-li/RISE-Forecasting}}.

\begin{figure}
    \centering
    \includegraphics[width=1.0\textwidth]{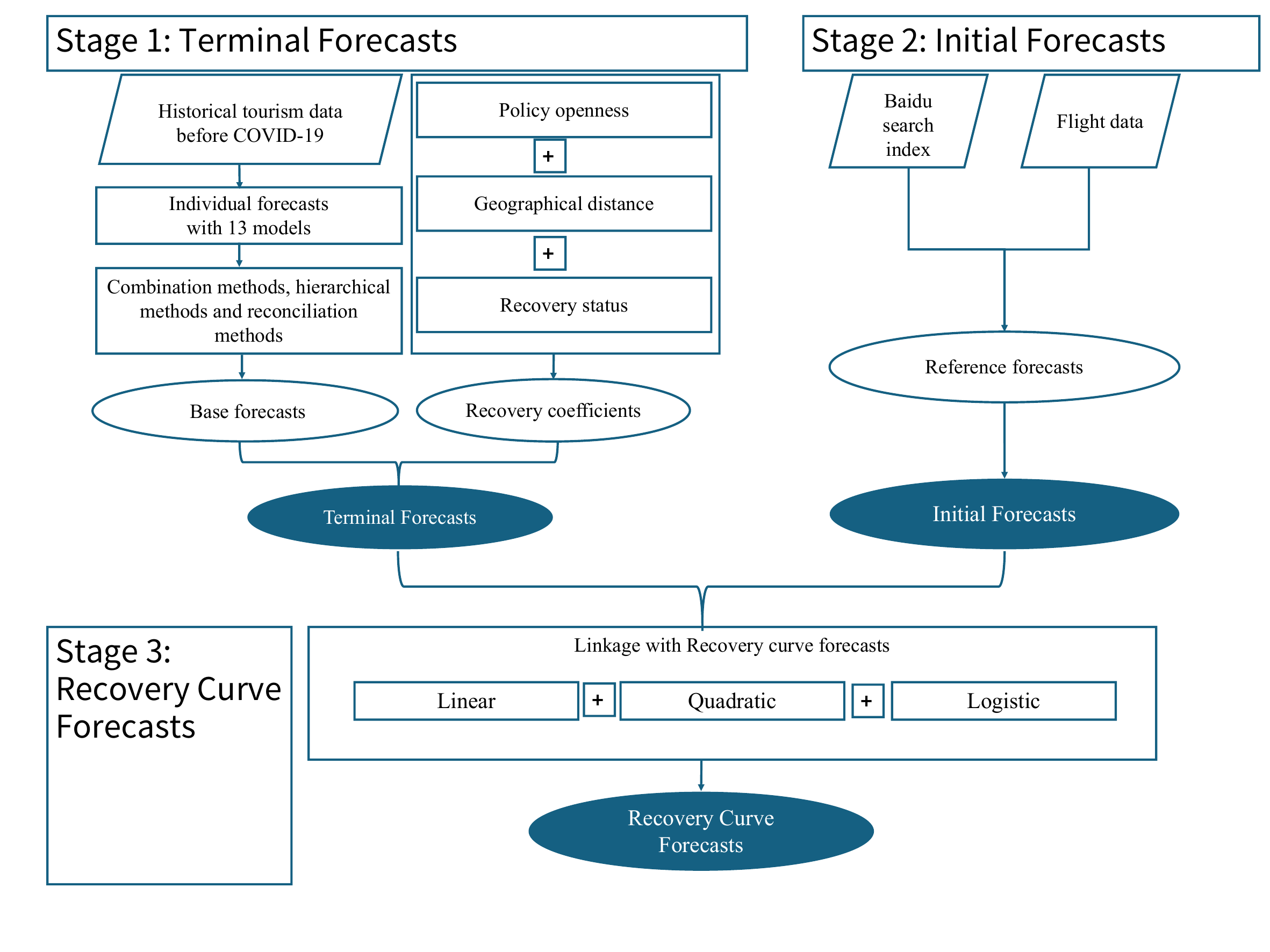}
    \caption{Flowchart of the Recovery-Informed Strategy Enhancement (RISE) framework.}
    \label{fig: flowchart}
\end{figure}

\begin{figure}
  \centering
    \includegraphics[width=1.0\textwidth]{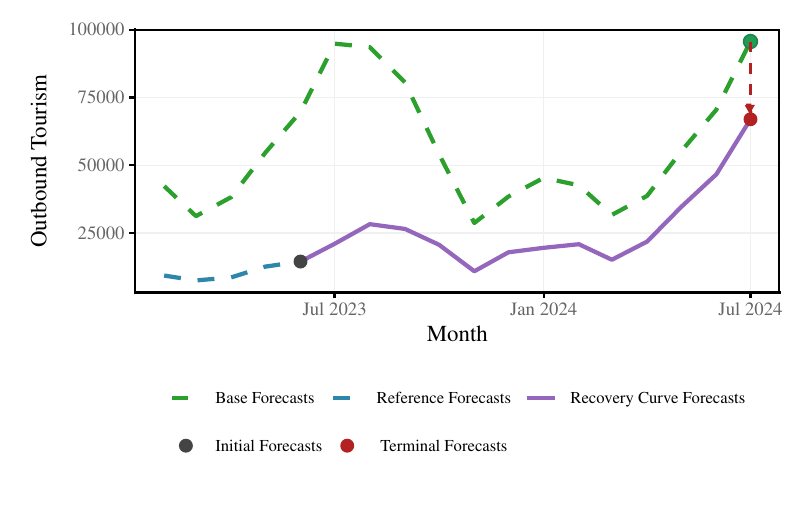}
    \caption{A sketch map of Recovery-Informed Strategy Enhancement (RISE) framework.}
    \label{fig: RISE}
\end{figure}

\subsection{First stage -- terminal forecasts}
\label{sec: base-forecast}

In the RISE framework, the terminal forecasts represent the end point in the forecast horizon, namely the level that the outbound tourism may recover to. As a result, to simulate the potential scenario of international tourism from Mainland China without the impact of COVID-19, we utilize 13 distinct forecasting models alongside three forecasting combination methods to create base tourism forecasts from 2023 to 2024 which are summarized in Table \ref{tab: models}. The time series data prior to 2018 serves as the training set, while data from 2018 to 2019 is employed as the validation set. Individual models are developed using the training set and assessed against the validation set to identify the optimal parameters for forecast combinations. Subsequently, all models are retrained using a combination of the training and validation sets to produce forecasts for the subsequent two years, specifically the projections for international tourism from 2020 to 2021 in a COVID-19-free context. As a result, these forecasts can be adjusted to serve as base projections for the recovery of the Chinese outbound travel market post-pandemic, specifically for the period 2023–2024.

\begin{table}[h!]
    \centering
    \caption{Description of individual forecasting models and hierarchical forecasting methods.}
    \label{tab: models}
    \resizebox{\textwidth}{!}{
    \begin{tabular}{p{4.5cm}p{10cm}p{5cm}}
        \toprule
        Model & Specification & Parameter setting \\
        \midrule
      Seasonal Na\"ive & The forecast is equal to the last observed value from the same season. & \texttt{snaive()} \\ \cline{2-3}
      Autoregressive Integrated Moving Average & The lags $(p, ~ d, ~ q)$ are automatically determined by \citet{HyndmanRJ2008AutomaticTime}. & \texttt{auto.arima()}\\ \cline{2-3}
      Random Walk with Drift & Equivalent to an ARIMA(0,1,0) model with an optional drift coefficient. & \texttt{rwf(drift=True)} \\ \cline{2-3}
      Exponential Smoothing State Space Model & State space family of models in \citet{HyndmanRJ2002StateSpace} & \texttt{ets(alpha=NULL, beta=NULL)} \\ \cline{2-3}
      Holt's Linear Trend & Exponential smoothing with a trend \citep{HoltCC1957ForecastingTrends}. & \texttt{holt(alpha=NULL, beta=NULL)} \\ \cline{2-3}
      Holt-Winters Seasonal & Extension of Holt's method to capture seasonality \citep{HoltCC1957ForecastingTrends,WintersPR1960ForecastingSales}       & \texttt{hw(alpha=NULL, beta=NULL)}      \\ \cline{2-3}
      Seasonal and Trend Decomposition with Loess & The trend component is estimated via ARIMA model and the seasonal component is estimated by & \texttt{mstl(iterate=2,s.window=NULL)} \\
              & (A) the average seasonal components of last three years, & \\
              & (B) the seasonal component of last year, and & \\
              & (C) estimated by ARIMA. & \\ \cline{2-3}
      State Space Trend–seasonal ARMA Model & State space model with Box-Cox transformation, ARMA errors, trend and seasonal components \citep{DeLiveraAM2011ForecastingTime}.   & \texttt{tbats(use.trend=NULL, seasonal.periods=NULL)} \\ \cline{2-3}
      Neural Network Autoregression & Feed-forward neural networks with one hidden layer and lagged values of the time series used as inputs. & \texttt{nnetar(p=12,P=1,size=7)} \\ \cline{2-3}
      Hierarchical Methods & TopDown hierarchical forecasting method that distributes the top level forecasts down hierarchies \citep{GrossCW1990DisaggregationMethods}. In method names, the letter A/B represents the proportion allocation approach determined by (A) the individual forecasts and (B) the history outbound tourism.  & \texttt{forecast(method="tdfp", fmethod="arima")} \newline \texttt{forecast(method="tdfp", fmethod="ets")} \newline  \texttt{forecast(method="tdgsa", fmethod="arima")} \newline \texttt{forecast(method="tdgsa", fmethod="ets")} \\ \cline{2-3}
      Reconciliation Methods & Minimum trace reconciliation method which minimizes the total forecast error while ensuring aggregation coherence. & \texttt{forecast(method = "comb", weights = "mint")} \\
              && \\
              & Weighted Least Squares method that assigns weights to different levels of the hierarchy to minimize reconciliation errors & \texttt{forecast(method = "comb", weights = "wls")} \\
  \bottomrule
 \end{tabular}
 }
\end{table}

All individual models used in this paper and their parameter settings are listed in Table \ref{tab: models}. The training process of 15 algorithms is performed in \textsf{R}. The methods are estimated by corresponding functions from the \textsf{R} package \textsf{forecast} \citep{HyndmanR2023ForecastForecasting}. All the hierarchical forecasting methods are performed with functions from the \textsf{R} packages \textsf{hts} \citep{HyndmanR2024HtsHierarchical}. The missing values in time series are imputed with Kalman smoothing method using function \texttt{na\_kalman()} in the \textsf{R} package \textsf{imputeTS} \citep{MoritzS2017ImputeTSTime}.

\subsubsection{Univariate forecast methods}

Our collection of models features a wide array of forecasting methods, including autoregressive techniques, exponential smoothing approaches, decomposition strategies, neural networks, and hierarchical forecasting models. Among all the models, two serve as na\"ive benchmarks: the seasonal na\"ive and random walk with drift models. Furthermore, we include the commonly used autoregressive integrated moving average (ARIMA) model, along with three variations of exponential smoothing models: simple exponential smoothing, Holt’s linear trend model, and the Holt-Winters seasonal model.

To effectively capture complex seasonal patterns and trends, we include three models: one based on the seasonal and trend decomposition with Loess method \citep{ClevelandRB1990STLSeasonaltrend}, and another using the exponential smoothing state space model with Box-Cox transformation, autoregressive moving average errors, and both trend and seasonal components. Additionally, we incorporate the neural network autoregressive model, which utilizes time series lag values as inputs to forecast future values, thereby improving the model's capacity to recognize intricate temporal dependencies.

\subsubsection{Forecast reconciliation methods}

\begin{figure}
    \centering
    \includegraphics[width=1.0\textwidth, trim={3cm 3cm 3cm 3cm}, clip]{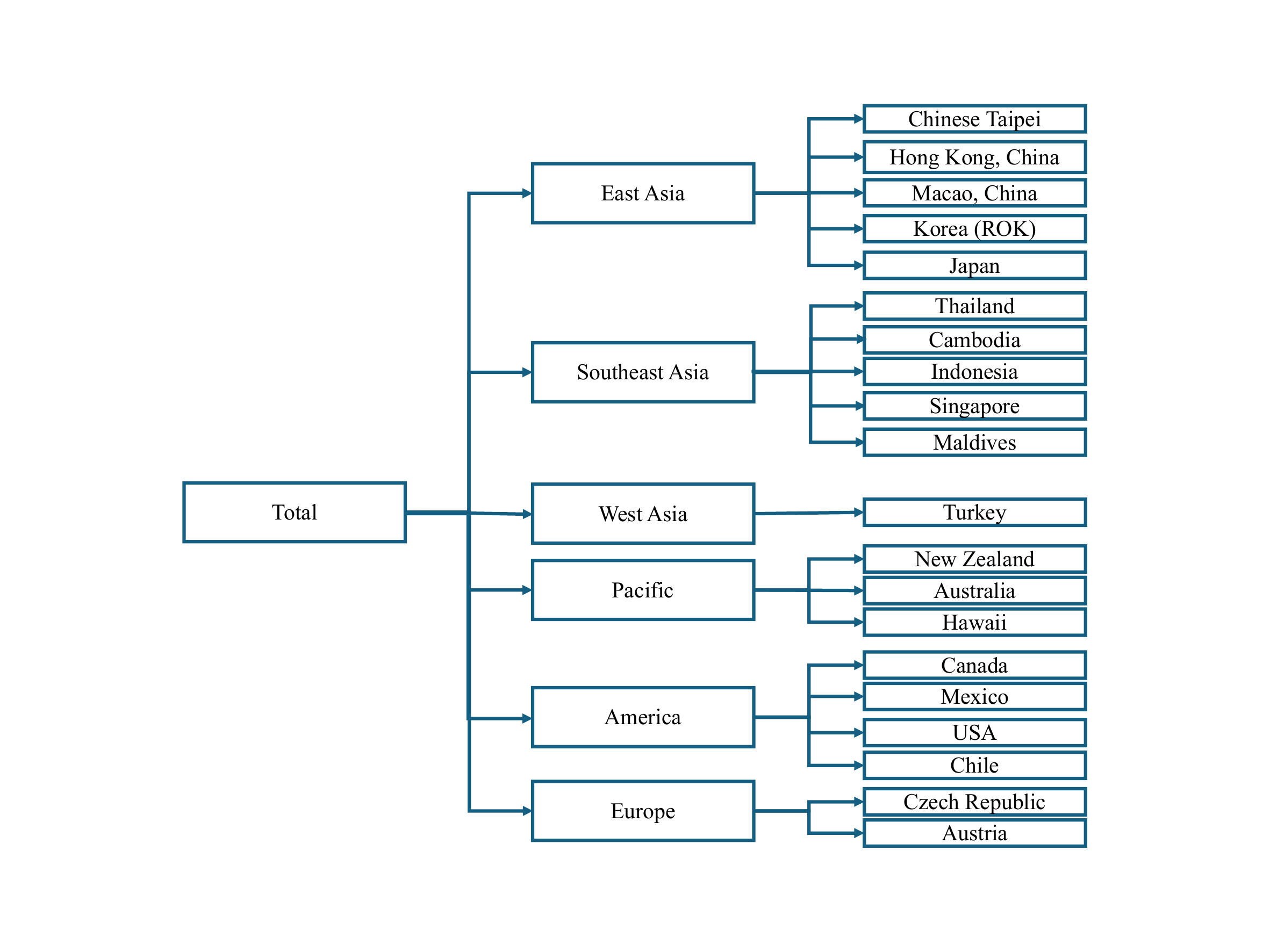}
    \caption{Hierarchical structure by region.}
    \label{fig: hierarchical}
\end{figure}

The aforementioned independent forecasts do not account for spatial correlation. The tourism data from the 20 destinations naturally form a three-level geographic hierarchical structure based on regions, which can enhance the accuracy of forecasts. We categorize the 20 destinations into six groups: East Asia, Southeast Asia, West Asia, Pacific, America and Europe, as illustrated in Figure \ref{fig: hierarchical}. We employ a hierarchical forecasting methodology, widely recognized in the forecasting community, see e.g. the M5 forecasting competition \citep{AndererM2022HierarchicalForecasting} and the tourism forecasting applications \citep{WenQ2026CoherentForecasts}; for a comprehensive overview, see \citet{AthanasopoulosG2024ForecastReconciliation}. Our approach incorporates two main strategies in hierarchical forecasting: Top-Down and Optimal Reconciliation.

Forecast reconciliation is a method used to ensure that forecasts made at different levels of a hierarchy (e.g., regions, departments, or products) are coherent — meaning they add up correctly. Let $\mathbf{y}_t$ be the vector of all time series at time $t$ (both aggregate and bottom-level),  $\widehat{\mathbf{y}}_t$ be base forecasts (possibly incoherent), $\widehat{\mathbf{y}}_t^{\text{rec}}$ reconciled (coherent) forecasts. The reconciled forecasts are given by
\begin{equation}
\widehat{\mathbf{y}}_t^{\text{rec}} = \mathbf{S} \mathbf{G} \widehat{\mathbf{y}}_t
\end{equation}
where  $\mathbf{S}$ is the summing matrix representing the hierarchical structure in Figure~\ref{fig: hierarchical} as
\begin{equation}
  \renewcommand{\arraystretch}{0.7}
  \setcounter{MaxMatrixCols}{20}
\scalebox{1}{$
\mathbf{S} =
  \begin{array}{cccccccccccccccccccc}
    \begin{bmatrix}
      \\
    1 & 1 & 1 & 1 & 1 & 1 & 1 & 1 & 1 & 1 & 1 & 1 & 1 & 1 & 1 & 1 & 1 & 1 & 1 & 1\\
    1 & 1 & 1 & 1 & 1 & 0 & 0 & 0 & 0 & 0 & 0 & 0 & 0 & 0 & 0 & 0 & 0 & 0 & 0 & 0\\
    0 & 0 & 0 & 0 & 0 & 1 & 1 & 1 & 1 & 1 & 0 & 0 & 0 & 0 & 0 & 0 & 0 & 0 & 0 & 0\\
    0 & 0 & 0 & 0 & 0 & 0 & 0 & 0 & 0 & 0 & 1 & 0 & 0 & 1 & 0 & 0 & 0 & 0 & 0 & 0\\
    0 & 0 & 0 & 0 & 0 & 0 & 0 & 0 & 0 & 0 & 0 & 1 & 1 & 1 & 0 & 0 & 0 & 0 & 0 & 0\\
    0 & 0 & 0 & 0 & 0 & 0 & 0 & 0 & 0 & 0 & 0 & 0 & 0 & 0 & 1 & 1 & 1 & 1 & 0 & 0\\
    0 & 0 & 0 & 0 & 0 & 0 & 0 & 0 & 0 & 0 & 0 & 0 & 0 & 0 & 0 & 0 & 0 & 0 & 1 & 1\\
    \multicolumn{20}{c}{\mathbf{I}_{m}}\\
  \end{bmatrix}
  \end{array}
  $}
\end{equation}
and $\mathbf{I}_{m}$ is an identity matrix,  $\mathbf{G}$ is a matrix that maps base forecasts to coherent forecasts. Different methods (like GLS and  minimum trace) estimate different $\mathbf{G}$.

Assume $\widehat{y}_{\text{total}, t}$ is the forecasts for the total series. The total is distributed to $m$ bottom-level series using proportions $\mathbf{p} = [p_1, p_2, \dots, p_m]^T$, where $\sum p_i = 1$. Suppose there are $n$ total series (including aggregates), and $m$ bottom-level series. The Top-Down approach, where forecasts are first made at the highest level of the hierarchy and then allocated to lower levels using proportions estimated via history information or individual forecasts.  Then $\mathbf{G}$ is $m \times n$ matrix
\begin{equation}
\mathbf{G} = \begin{bmatrix}
p_1 & 0 & \cdots & 0 \\
p_2 & 0 & \cdots & 0 \\
\vdots & \vdots & \ddots & \vdots \\
p_m & 0 & \cdots & 0 \\
\end{bmatrix}
\end{equation}
where $n$ is the number of nodes in the whole hierarchy. In our case, $n = 27$.

The optimal reconciliation method \citep{WickramasuriyaSL2019OptimalForecast} applies a reconciliation step using statistical techniques such as Weighted Least Squares or Minimum Trace to adjust individual forecasts optimally. The minimum trace (MinT) method sets $\mathbf{G}$ to minimize the trace of the covariance matrix of reconciliation errors. The solution is
\begin{equation}
\mathbf{G} = \left( \mathbf{S}^\top \mathbf{W}^{-1} \mathbf{S} \right)^{-1} \mathbf{S}^\top \mathbf{W}^{-1}
\end{equation}
where $\mathbf{W}$ is the covariance matrix of the base forecasts errors. This gives a generalized least squares solution under the constraint that the reconciled forecasts lie in the coherent subspace defined by $\mathbf{S}$.

\subsubsection{Forecast combination methods}
Three distinct forecast combination methods are utilized to aggregate the individual base forecasts: equally weighted average, error-based weighted average, and stacking models. Numerous studies have indicated that the simple arithmetic average of basic forecasts can achieve performance comparable to more complex forecasting combination methods \citep{WangX2023ForecastCombinations}. In the error-based weighted average method, data prior to 2018 serve as the training set, while data from 2018 and 2019 are used as the validation set. The weight assigned to each base model is calculated as the inverse of the Mean Absolute Percentage Error of the validation set. For the stacking methods, Lasso and Ridge regression are selected as the meta models. In both approaches, the regression coefficients are constrained to be strictly positive without including the intercept. The regularization parameter is set as $1$ in both models.

We choose the top $80\%$ of models based on accuracy to participate in the forecast combination procedure for generating base forecasts. Consequently, $4$ out of the $17$ models are excluded due to their significantly higher forecast error in the validation set compared to the others. As a result, $13$ individual models are included in the forecast combination step. Detailed information on errors and models for the base forecasts can be found in Table \ref{tab:table2}. The base forecasts in month $t$ is denoted as $\widehat{BF}_i^{t}$.

\subsubsection{Recovery coefficient}

Although China lifted outbound travel restrictions at the end of 2022, the impact of the pandemic on the tourism market has not been immediately eliminated. Historical data can no longer accurately reflect post-pandemic tourism demand, and relying solely on traditional time series models might still result in considerable prediction bias. Since the pace of tourism recovery varies across countries due to factors such as pandemic policies, flight availability, and travel costs, we introduce a recovery coefficient $r_i$ where $0<=r_i<=1$ for each destination and define the terminal forecasts as $\widehat{BF}_i^{202407}*r_i$

\citet{SkareM2021ImpactCOVID19} pointed out that the recovery of tourism industry worldwide would take more time than the average expected recovery period of 10 months amid COVID-19. Since China has imposed stricter travel restrictions than most of other destinations, and considering the fact that the number of outbound flights has not yet returned to pre-pandemic levels, it might take much time for travelers to regain their enthusiasm and habitual engagement in international travel. Consequently, as of August 2024, 18 months after the end of tourism restrictions, it is unlikely for China's outbound tourism to completely recover from the impact of COVID-19 and the base forecasts in August 2024 must be adjusted to account for such influence.

An ideal way to estimate the recovery coefficient is to learn it from data. However, in the forecasting competition, the actual value ${F}$ is unknown. Therefore, we treat the recovery coefficient as a hyperparameter whose value is obtained from prior knowledge. \citet{LiX2021ImpactsCOVID19} pointed out that travelers' preferences have shifted towards nearby and low-risk destinations post-COVID-19. In the aftermath of the COVID-19 pandemic, travelers may favor shorter trips due to health concerns, travel restrictions, and economic uncertainty. The tourism policy is also a critical factor for tourism departures \citep{NeumayerE2010VisaRestrictions}.

As a result, we use policy information, distance and the recovery situation as the main prior reference for setting the recovery coefficient. For each destination, policy, distance, and recovery score are assigned on a scale from 1 to 5. The policy and recovery scores in Table \ref{tab:recovery-coef} were determined based on observable indicators reflecting each destination’s border control, visa policy, and pace of tourism reopening. Following \citet{ZhangH2021ForecastingTourism}, destinations with close geographical proximity and strong cross-border linkages, such as Hong Kong and Macao, tend to recover more rapidly than long-haul markets like the USA. Since December 2022, most East and Southeast Asian destinations have gradually reopened to Chinese travelers, reinstating direct flights and easing entry restrictions. Their high policy and recovery scores reflect early resumption of air services, relatively low travel costs, and supportive government attitudes toward inbound tourism. Chinese Taipei received the lowest policy score because the July 2019 suspension of individual travel permits from Mainland China remains in effect, limiting outbound tourism recovery despite regional reopening. European destinations and the USA were assigned lower policy and recovery scores due to stricter visa requirements, slower normalization of flight capacity, and longer travel distances, all of which hinder short-term recovery. Chile received a moderate recovery score, taking into account emerging bilateral trade and air-route expansion. For Pacific destinations such as Australia, Hawaii, and Turkey, moderate policy openness and partial restoration of flights support mid-range recovery scores, while New Zealand was rated slightly lower because of its conservative border management and slower recovery of long-haul connectivity.

The distance score is determined by the average distance from Beijing and Shanghai to the capital city of the destination. The recovery score is determined by the forecasts of tourism arrivals and the latest flight data. The policy score is evaluated with the destination’s travel restriction policies, the difficulty of visa applications, and the governments' attitude toward Chinese tourists. Their values are listed in Table \ref{tab:recovery-coef} and more detailed information can be found in supplementary materials in the GitHub repository.

\begin{table}[htbp]
  \centering
  \caption{Evaluation scores and recovery coefficients for each destination. }
  \label{tab:recovery-coef}
  \resizebox{\textwidth}{!}{
  \begin{tabular}{cccccccc}
    \toprule
    Region & Destination      & Policy & Distance & Recovery & Average      & Recovery coefficient \\
    \midrule
    \multirow{4}{*}{America}
           & Canada           & 3      & 1        & 2        & \textbf{2.0} & \textbf{0.65}        \\
           & Chile            & 3      & 1        & 3        & 2.3          & 0.7                  \\
           & Mexico           & 5      & 1        & 5        & \textbf{3.7} & \textbf{1.0}         \\
           & USA              & 2      & 1        & 3        & 2.0          & 0.65                 \\ \cline{1-7}
    \multirow{5}{*}{East Asia}
           & Chinese Taipei   & 1      & 5        & 1        & 2.3          & 0.6                  \\
           & Hong Kong, China & 5      & 5        & 3        & \textbf{4.3} & \textbf{0.85}        \\
           & Macao, China     & 5      & 5        & 3        & 4.3          & 0.85                 \\
           & Korea (ROK)      & 4      & 5        & 2        & 3.7          & 0.8                  \\
           & Japan            & 4      & 5        & 2        & 3.7          & 0.8                  \\ \cline{1-7}
    \multirow{5}{*}{Southeast Asia}
           & Thailand         & 5      & 3        & 4        & 4.0          & 0.8                  \\
           & Cambodia         & 5      & 3        & 3        & 3.7          & 0.8                  \\
           & Indonesia        & 4      & 3        & 4        & 3.7          & 0.8                  \\
           & Singapore        & 4      & 3        & 3        & 3.3          & 0.8                  \\
           & Maldives         & 4      & 3        & 5        & 3.7          & 0.8                  \\ \cline{1-7}
    \multirow{3}{*}{Pacific}
           & New Zealand      & 3      & 1        & 3        & 2.3          & 0.7                  \\
           & Australia        & 4      & 2        & 3        & 3.0          & 0.75                 \\
           & Hawaii           & 3      & 2        & 4        & 3.0          & 0.75                 \\ \cline{1-7}
    \multirow{1}{*}{West Asia}
           & Turkey           & 4      & 2        & 3        & 3.0          & 0.75                 \\ \cline{1-7}
    \multirow{2}{*}{Europe}
           & Austria          & 2      & 2        & 2        & 2.0          & 0.65                 \\
           & Czech Republic   & 2      & 2        & 2        & 2.0          & 0.65                 \\
    \bottomrule
  \end{tabular}
}
\end{table}

The recovery coefficient ($r_i$) for each destination, as shown in Table \ref{tab:recovery-coef} is determined by the average of three standardized indicators, policy openness, geographical distance, and recovery status. These indicators capture, respectively, institutional, spatial, and market-driven constraints on tourism recovery. We assume that $r_i$ increases linearly with the average score, reflecting that destinations with more open policies, shorter distances, and stronger recovery signals tend to experience faster normalization of tourism flows. To calibrate the mapping, we assign the recovery coefficients for three destinations with well-documented recovery patterns Canada ($r=0.65$), Mexico ($r=1.0$), and Hong Kong, China ($r=0.85$) as anchor points. Canada is assigned a coefficient of 0.65, mainly due to stringent visa policies for Chinese citizens, incomplete restoration of flight capacity, and longer travel distances. Mexico is granted a recovery coefficient of 1.0, largely because of its significant tourism growth from China. By December 2022, China's outbound tourism to Mexico had rebounded to 85\% of pre-pandemic levels. The recovery coefficient for Hong Kong is set at 0.85, reflecting its close ties to Mainland China and the convenient entry arrangements for Chinese residents. A least-squares regression based on these reference values yields
$r_i = 0.45 + 0.10 \times \text{Average}_i$. The remaining coefficients were then generated using this fitted model. This procedure ensures internal consistency across destinations and links the assigned coefficients directly to measurable recovery determinants rather than subjective judgment.

By combining a quantitative scoring system with expert evaluation, we ensure cross-destination consistency while capturing policy dynamics and market conditions that are inherently difficult to model statistically. Human judgment is explicitly embedded in the estimation of recovery coefficients, allowing the framework to incorporate contextual and forward-looking insights beyond what historical data alone can provide. This hybrid design improves the realism and adaptability of recovery forecasts, especially in post-crisis environments characterized by high uncertainty and limited data availability.

\subsection{Second stage -- initial forecasts}

The second stage of the forecasting strategy focuses on estimating the reference forecasts by incorporating external factors, such as search index data and flight information. This phase is critical for improving the accuracy of initial forecasts.

The competition's forecast period runs from August 2023 to July 2024, and the latest available historical tourism data only extends until January 2023. Short-term forecasting of the outbound tourism in the gap period from February to July 2023 using relevant external indicators can make a significant contribution to long-term forecasting accuracy. In accordance with the submission deadline, we only forecast the data from February 2023 to June 2023, which is referred to as the \emph{reference forecasts} in the subsequent discussion. This approach allows us to integrate the most recent external information (recovery information) into our framework. The external factors considered in this stage include the Baidu composite search index and the number of nonstop flights between Mainland China and 20 key tourist destinations.

\subsubsection{Baidu search index}

The Baidu Index website has provided daily search index data for various keywords since 2010 and is widely used for tourism forecasting because it provides rapid information \citep{LiS2018EffectiveTourist}. For each destination, we create a Baidu composite search index by aggregating multiple keywords related to tourist arrivals, including terms such as travel, shopping, food, hotels, visas, and the names of key cities and attractions. Due to differences in data availability and regional conditions, the specific keywords used to formulate the Baidu composite search index vary across destinations. A comprehensive list of these keywords is available in Table~\ref{tab: baidu_keywords}. The correlation between the search indices of each keyword and the corresponding outbound tourism is calculated, and keywords with a correlation coefficient less than 0.6 are excluded from the construction of the composite Baidu Index.

\begin{table}
 \caption{Baidu Search Index keywords by destination.}
 \label{tab: baidu_keywords}
 \begin{tabular}{lp{12cm}}
  \toprule
   Destination   & Keywords             \\
    \midrule
    Canada           & Canada travel, Canada visa, Air Canada, Vancouver                                                    \\
    Chile            & Chile travel, Chile visa, Santiago                                                                                                                                                                                           \\
    Mexico           & Mexico travel, Mexico visa                                                                                                                                                                                                   \\
    USA              & USA travel, USA travel guide, USA airlines, USA visa, Los Angeles, San Francisco                                                                                                                                         \\
    Chinese Taipei   & Chinese Taipei travel, Chinese Taipei travel guide, Chinese Taipei hotel, Chinese Taipei visa, Chinese Taipei cuisine, Chinese Taipei snacks, Chinese Taipei shopping                                                                                                                \\
    Hong Kong, China & Hong Kong travel guide, Hong Kong airlines, Hong Kong hotel, Hong Kong visa, Hong Kong cuisine, Hong Kong snacks, Hong Kong shopping, Hong Kong travel map, Hong Kong tourist attractions, Hong Kong and Macao travel permit \\
    Macao, China     & Macao travel, Macao travel guide, Macao airlines, Macao hotel, Macao cuisine, Hong Kong and Macao travel permit                                                                                                              \\
    Korea (ROK)      & Korea travel, Korea travel guide, Korea hotel, Korea visa, Korea cuisine, Korea shopping, Korea tourist attractions, Seoul, Jeju Island                                            \\
    Japan            & Japan travel, Japan travel guide, Japan airlines, Japan visa, Japanese cuisine, Japan shopping, Japan travel map, Japan tourist attractions, Osaka, Nagoya                                                                   \\
    Thailand         & Thailand travel, Thailand airlines, Thailand hotel, Thailand visa, Bangkok, Chiang Mai, Koh Samui, Phuket                                                                                                                    \\
    Cambodia         & Cambodia travel, Cambodia visa, Angkor Wat                                                                                                                                                                                   \\
    Indonesia        & Indonesia travel, Indonesia visa, Bali                                                                                                                                                                                       \\
    Singapore        & Singapore travel, Singapore travel guide, Singapore airlines, Singapore visa                                                                                                                                                 \\
    Maldives         & Maldives travel, Maldives travel guide, Maldives visa                                                                                                                                                                        \\
    New Zealand      & New Zealand travel, New Zealand travel guide, New Zealand airlines, New Zealand visa                                                                                                                                         \\
    Australia        & Australia travel guide, Australia visa, Sydney, Melbourne                                                                                                                                                                    \\
    Hawaii           & Hawaii travel, Hawaii travel guide                                                                                                                                                                                           \\
    Turkey           & Turkey travel, Turkey travel guide, Turkish airlines, Turkey hotel, Turkey visa                                                                                                                                              \\
    Austria          & Austria travel, Austrian airlines, Austria visa                                                                                                                                                                              \\
    Czech Republic   & Czech Republic travel, Czech Republic visa                                                                                                                                                                                   \\
    \bottomrule
  \end{tabular}
\end{table}

\begin{figure}
    \centering
    \hspace*{-1.3cm}
    \includegraphics[scale=1.0]{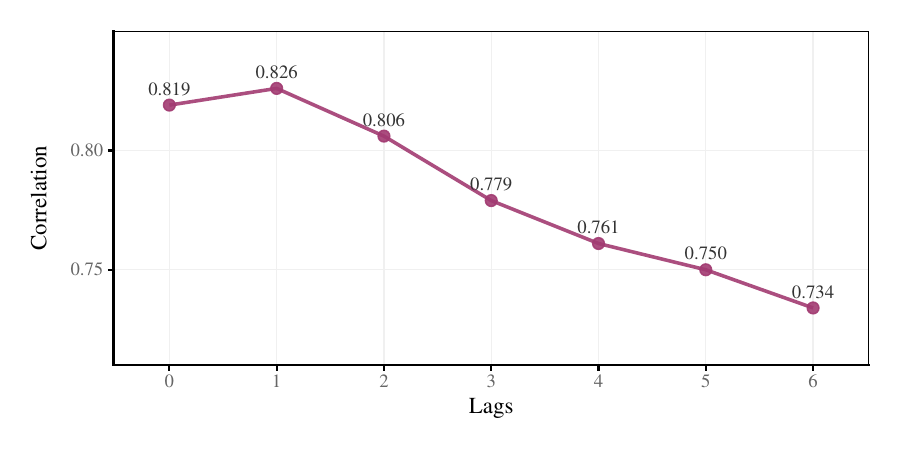}
    \caption{Correlation between Baidu Search Index and outbound tourism at different time lags.}
    \label{fig: lags_correlation}
\end{figure}

\begin{figure}
    \centering
    \includegraphics[width=1.0\textwidth]{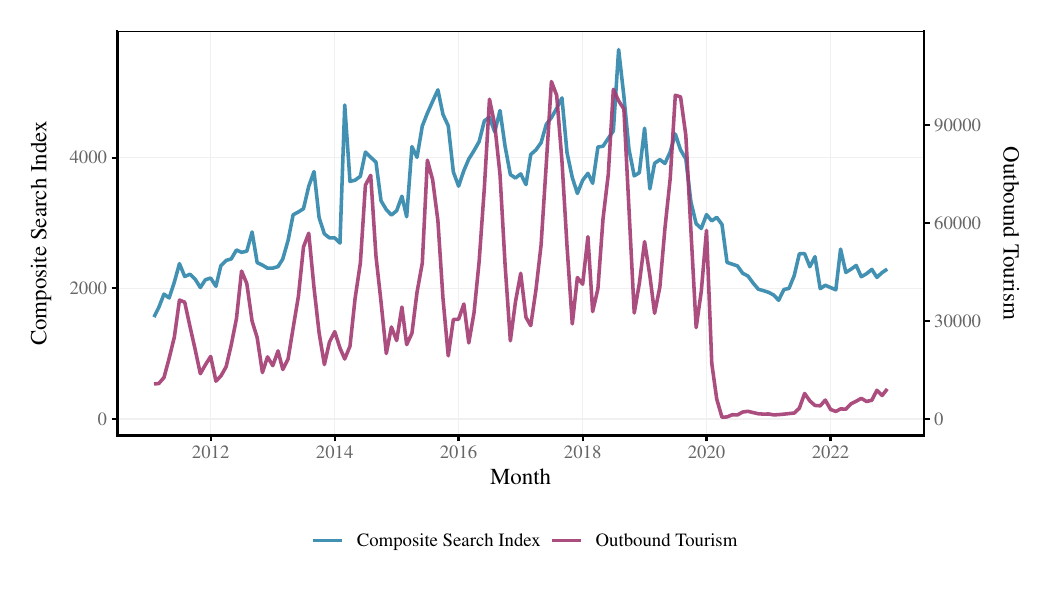}
    \caption{The relationship between composite search index (in blue with left axis) and outbound tourism (in red with right axis) to Canada.}
    \label{fig: search_index}
\end{figure}

The primary data source for outbound tourism, a monthly cross-sectional dataset, is provided by the organizing committee of \emph{Tourism Forecasting Competition amid COVID-19 Round II}. We gathered search data, which was aggregated monthly to align with the outbound tourism data. A one-month lagged Baidu search index is used as a predictor for current outbound tourism, reflecting tourists’ tendency to search for travel information in advance. Figure \ref{fig: lags_correlation} illustrates the average correlation coefficients between outbound tourism and the Baidu search index at various time lags, with the highest correlation (0.826) observed at a one-month lag. Additionally, we calculated the correlation coefficients between individual keywords and outbound tourism; keywords with coefficients below 0.6 were excluded from the composite search index construction. The Baidu composite search index was then obtained by summing the remaining keyword indices. As shown in Figure \ref{fig: search_index}, the constructed composite search index exhibits a strong correlation with outbound tourism.

Two primary strategies are utilized to estimate the reference forecasts using the Baidu composite search index. The first strategy employs the Baidu composite search index as an external variable within two multivariate time series models: ARIMA with exogenous variables and Facebook Prophet model \citep{TaylorSJ2018ForecastingScale}. The second strategy involves calculating the ratio of outbound tourism to the Baidu composite search index and applying several univariate time series models, including Exponential Smoothing \citep{HyndmanRJ2002StateSpace}, Holt-Winters, and the state-space trend–seasonal ARMA model \citep{DeLiveraAM2011ForecastingTime}, to predict this ratio. The estimate of the reference forecasts in this approach is obtained by multiplying the predicted ratio by the Baidu composite search index. The results obtained from the above strategies are then averaged.

\subsubsection{Flight information}
Flight data offer essential supplementary information for estimating outbound tourism trends. The VariFlight app provides flight data from 2019 to 2023, detailing the number of direct flights between two destinations. Initially, we aimed to estimate the ratio of actual outbound tourists, as provided by the competition organizer, to the number of direct flights, following a methodology similar to the Baidu index. However, due to air traffic control policies implemented over the past three years, this ratio has decreased significantly, making predictions based on it unreliable. For similar reasons, multivariate time series models are also impractical. Therefore, a simple predictive approach is adopted: we start with the most recent outbound tourism data as the baseline. Then, for a given month, we compute the relative growth rate of flights compared to the initial month. The forecasted outbound tourism value is obtained by multiplying the baseline number by this growth rate.

The reference forecasts in month $t$ are denoted as $\widehat{RF}_i^{t}$ as the simple average of the reference forecast based on Baidu composite search index and flight data. Similarly, the initial forecasts can be extracted from reference forecasts as $\widehat{RF}_i^{202306}$

\subsection{Third stage -- recovery curve forecasts}

\subsubsection{Recovery curve estimation}

In this section, the recovery curve forecasts that connect the initial forecasts (June 2023) and the terminal forecasts (July 2024) are estimated using three different functional forms-linear, quadratic, and logistic—to capture a range of possible recovery paths and ensure robust forecasting under uncertainty.

First, we define the monthly forecasts for outbound tourism from Mainland China to destination $i$ as ${F}_i^t$, where $t=0,1,\dots,13$, corresponds to month from June 2023 to July 2024. More specifically, ${F}_i^0$ represents the forecasts for June 2023, ${F}_i^1$ for July 2023, and so on, with ${F}_i^{13}$ denoting the forecasts for July 2024.

Two critical points, the initial forecasts and the terminal forecasts play an important role in our forecast framework. As we have mentioned before, the initial forecast ${F}_i^0$ is set as the reference forecasts in June 2023, and the terminal forecasts ${F}_i^{13}$ are the multiplication of the recovery coefficient and the base forecasts in July 2024. They are denoted as:
\begin{equation}
\label{eq: initial-forecasts}
{F}_i^{0} = \widehat{RF}_i^{202306}
\end{equation}

\begin{equation}
\label{eq: terminal-forecasts}
{F}_i^{13} = \widehat{BF}_i^{202407} \times r_i
\end{equation}
where $0 < r_i \leq 1$.

In terms of recovery curve estimation, we employ the seasonal and trend factorization in its multiplicative form of the trend and seasonality to derive the final recovery curve, namely
\begin{equation}
\label{eq: stl}
{F}_i^{t} = \widehat{Trend}_i^{t}\times\widehat{Seasonal}_i^{t}.
\end{equation}
Below, we estimate the trend and seasonal components involved in the seasonal and trend factorization.

The seasonal component $\widehat{Seasonal}_i^{t}$ is estimated by applying the seasonal and trend decomposition in its multiplicative form to the logarithm of the historical outbound tourism series prior to 2020. This logarithmic series facilitates a multiplicative seasonal adjustment that is better suited for recovery forecasting. Note that the logarithmic series is always positive because we use the original scale. While alternative methods for estimating the seasonal component exist -- such as using base forecasts post-August 2023 -- we contend that the seasonal effects are more stable within historical data, and relying on base forecasts may introduce unwarranted noise.

In this study, we propose that the trend components of the recovery curve can adopt one of three functional forms: linear, quadratic, or logistic. We utilize these functions to fit the trend component accordingly.

For the linear trend estimation, we establish a straight line linking the initial and terminal forecasts. The trend component estimate for destination $i$ in month $t$ based on linear model can be expressed as:

\begin{equation}
\label{eq:linear_trend}
\widehat{Trend}_i^t = \widetilde{Trend}_i^0 + \frac{t}{14} \cdot (\widetilde{Trend}_i^{13} - \widetilde{Trend}_i^0), \quad t = 0, 1, \dots, 13,
\end{equation}
where
\begin{equation}
\label{eq: trend}
\widetilde{Trend}_i^{0} = \frac{{F}_i^{0}}{\widehat{Seasonal}_i^{0}}, \quad \text{and} ~
\widetilde{Trend}_i^{13} = \frac{{F}_i^{13}}{\widehat{Seasonal}_i^{13}}
\end{equation}
are the estimated intercepts for initial and terminal points. This is similar to \citet{QiuRTR2021VisitorArrivals} in which they construct a recovery curve solely assuming a linear relationship.

For the quadratic trend estimation, we fit a quadratic curve on the data from January 2022 to June 2023 and the terminal forecasts utilizing the Least Squares method. The data for the first 13 months (from January 2022 to January 2023) are actual values while the next five months (from February 2023 to June 2023) are reference forecasts. The implicit assumption of the quadratic curve is that the number of outbound tourism recovers slowly at first, but rises swiftly after a period of time. Notably, all the data are trend components and the weight of the terminal forecasts is set to $18$ to enhance the curve's fitness. The trend component estimate based on quadratic model can be expressed as:
\begin{equation}
\label{eq:quadratic_trend}
\widehat{Trend}_i^t = a_i(t+18)^2 + b_i(t+18) + c_i,
\end{equation}
for $t = 0, ~ 1, ~ \dots, 13$, where
\begin{equation}
a_i, ~ b_i, ~ c_i=\arg\min_{a_i, b_i, c_i} \big( \sum_{t=1}^{18} \left( y_i^t - (a_i t^2 + b_it + c) \right)^2 + 18\times(\widehat{Trend}_i^{13}- (a_i t^2 + b_i t + c) )^2 \big)
\end{equation}
The trend forecasts before February 2023 are extracted from historical data
\begin{equation}
y_i^1={Trend}_i^{202201} , ~ y_i^2={Trend}_i^{202202},...,y_i^{13}={Trend}_i^{202301}
\end{equation}
and the trend forecasts between February to 2023 to June 2023 is the reference forecasts, as
\begin{equation}
y_i^{14}=\frac{\widehat{RF}_i^{202302}}{\widehat{Seasonal}_i^{202302}}, ~
y_i^{15}=\frac{\widehat{RF}_i^{202303}}{\widehat{Seasonal}_i^{202303}},..., ~
y_i^{18}=\frac{\widehat{RF}_i^{202306}}{\widehat{Seasonal}_i^{202306}}
\end{equation}

In the case of the logistic trend estimation, the trend component of five critical points $y_1, ..., y_5$ are employed to estimate the logistic curve: the actual value in January 2022, the initial forecasts and the base forecasts in December 2023, July 2024 and December 2024. They are expressed as
\begin{equation}
y_i^1={Trend}_i^{202201} , ~
y_i^2={\widetilde{Trend}_i^{0}}, ~
y_i^3=\frac{{\widehat{BF}_i^{202312}}}{\widehat{Seasoal}_i^{202312}}, ~
y_i^4=\frac{{\widehat{BF}_i^{202407}}}{\widehat{Seasoal}_i^{202407}}, ~
y_i^5=\frac{{\widehat{BF}_i^{202412}}}{\widehat{Seasoal}_i^{202412}}.
\end{equation}
The estimation is carried out using the \textsf{R} package \textsf{nls}. The trend component estimate based on logistic model can be expressed as
\begin{equation}
\widehat{Trend}_i^t=\frac{L_i}{1 + \exp(-k_i(t +18 - t_i))},
\end{equation}
where
\begin{equation}
L_i, ~ k_i, ~ t_i=\arg\min_{L_i, k_i, t_i} \sum_{t=1}^{5} \left( y_i^t - \frac{L_i}{1 + \exp(-k_i(t - t_i))} \right)^2.
\end{equation}

\begin{figure}
  \centering
    \includegraphics[width=1.0\textwidth]{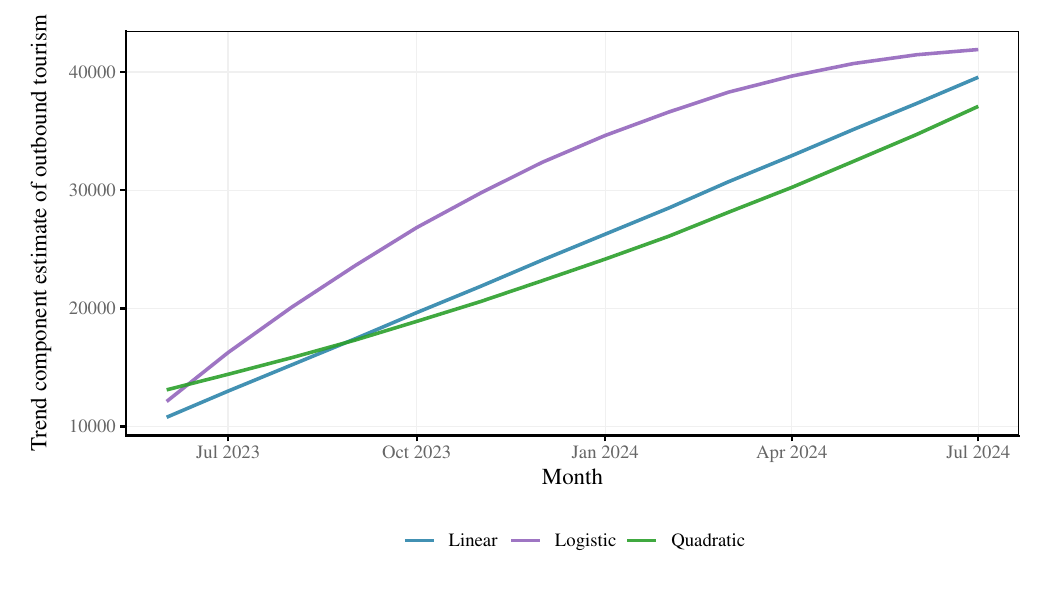}
    \caption{Three trend component functions in the recovery curve forecasting stage for outbound tourism to Canada.}
    \label{fig: recovery_curve_example}
\end{figure}

We demonstrate three types of recovery curves, as shown in Figure \ref{fig: recovery_curve_example} for the trend component forecasts of Canada. It is worth noting that the final submitted forecasts in July 2024 do not exactly match the terminal forecasts, but rather approximate them. In the quadratic and logistic approaches, the forecasts for July 2024 are determined by the specific curve fitted to the data, and the value may deviate slightly depending on the model parameters and fitting method.

Finally, we derive the trend component estimate by averaging the results generated by the three aforementioned methods. The recovery curve forecasts are the multiplication of the trend components estimate and the corresponding seasonal components estimate as Equation (\ref{eq: stl}).

\subsection{Interval forecast}

In the RISE framework, the forecasting process is divided into three parts: the initial forecast, the terminal forecast, and the recovery curve that connects them. Among the $13$ individual models described in Section~\ref{sec: base-forecast}, $12$ models (excluding the neural network autoregressive model, which does not produce prediction intervals) generate interval forecasts. For each month, we use the average of the 80\% upper and lower confidence bounds from these $12$ models as the corresponding bounds for the terminal forecast. Following the same procedure as in the point forecast, these upper and lower terminal forecasts are linked to the initial forecast through the recovery curve to construct the complete interval forecast. In the competition, the interval forecasts were evaluated using the Winkler score \citep{WinklerRL1972DecisionTheoreticApproach}, which measures both the width of the interval and the penalty for observations falling outside it as $Winkler_{\alpha}=\frac{1}{T}\sum_{t=1}^{T}W_{\alpha,t}$ and
\begin{equation}
\label{eq: winkler_total}
W_{\alpha,t} = \begin{cases}
  (u_{\alpha,t} - \ell_{\alpha,t}) + \frac{2}{\alpha} (\ell_{\alpha,t} - y_t) & \text{if } y_t < \ell_{\alpha,t}, \\
  (u_{\alpha,t} - \ell_{\alpha,t})   & \text{if }  \ell_{\alpha,t} \le y_t \le u_{\alpha,t}, \\
  (u_{\alpha,t} - \ell_{\alpha,t}) + \frac{2}{\alpha} (y_t - u_{\alpha,t}) & \text{if } y_t > u_{\alpha,t},
  \end{cases}
\end{equation}
where $u_{\alpha, t}$ and $l_{\alpha, t}$ are the corresponding upper and lower bounds of the $100(1 - \alpha)\%$ interval. To account for differences in scale across series, a standardized version of the Winkler score is also reported as
\begin{equation}
\label{eq: winkler_total_st}
StandardWinkler_{\alpha}=Winkler_{\alpha}\div\frac{\sum_{i=1}^{T}y_{\alpha,t}}{T}.
\end{equation}
In addition, the coverage probability is reported to evaluate how well the prediction intervals align with the nominal confidence level as
\begin{equation}
\label{eq: cp_total}
CP=\frac{1}{T}\sum_{t=1}^{T}CP_{t} ~\text{and} ~ CP_t = \begin{cases}
  0 & \text{if } y_t < \ell_{\alpha,t}, \\
  1   & \text{if }  \ell_{\alpha,t} \le y_t \le u_{\alpha,t}, \\
  0 & \text{if } y_t > u_{\alpha,t}.
  \end{cases}
 \end{equation}

\section{Forecast evaluation}
\label{sec: evaluation}

\subsection{Base forecasts}

Table \ref{tab:table2} presents the Mean Absolute Percentage Error and the Mean Absolute Scaled Error for the base forecasts in the validation set during 2018--2019 with all forecast models. Among all individual models, two hierarchical forecasting models with the weighted least squares and minimum trace reconciliation methods demonstrated one of the best performances. Exponential smoothing method also performed well. In contrast, Random walk with drift model, Holt’s linear trend model, TopDown with ARIMA and TopDown with ETS models where the distributing proportion is determined by historical outbound tourism are dropped for their unusually high validation error. Therefore, they are excluded from our model list. The results indicate that allocating proportions based on forecast values yields better performance than using historical data in the top-down method. Combination methods, especially stacking models, significantly decrease forecast error compared with individual models. The Stacking-Lasso method achieves the lowest validation error 0.0961 for Mean Absolute Percentage Error and 0.5421 for Mean Absolute Scaled Error. The results highlight the effectiveness of forecast combination, particularly stacking-based approaches, in improving forecast accuracy over individual models.

\begin{table}
    \centering
    \caption{Validation set for the base forecasts during 2018 -- 2019 for all models. }
    \label{tab:table2}
    \resizebox{\textwidth}{!}{
      \begin{tabular}{p{7cm}cccc}
        \toprule

        \multicolumn{3}{c}{Panel A: errors of individual models}                                 \\
        \midrule
        Model                             &Root Mean Squared Error                & Mean Absolute Percentage Error            & Mean Absolute Scaled Error            \\
        \midrule
        Seasonal Na\"ive                &94969                & 0.2679          & 1.4648          \\
        Random walk with drift       &\textbf{118231}           & \textbf{0.4158} & \textbf{1.8741} \\
        ARIMA      &107064       & 0.2430          & 1.2175          \\
        Exponential smoothing             &107330                   & 0.1494          & 0.9523          \\
        Holt's linear trend           &\textbf{143351}                       & \textbf{0.4350} & \textbf{3.3307} \\
        Holt-Winters seasonal       & 88134                         & 0.1851          & 1.1028          \\
        Seasonal and Trend Decomposition with Loess-A    &98743    & 0.1769          & 1.1512          \\
        Seasonal and Trend Decomposition with Loess-B   &107063     & 0.2249          & 1.2802          \\
        Seasonal and Trend Decomposition with Loess-C   &101330     & 0.1859          & 1.1830          \\
        TBATS                   &120628                             & 0.2796          & 1.4949          \\
        Neural Network Autoregressive     &103085                   & 0.1994          & 1.3337          \\
        TopDown-A-ARIMA                &101617                      & 0.2220          & 1.2520          \\
        TopDown-A-ETS                  &101593                      & 0.1550          & 0.9860          \\
        TopDown-B-ARIMA              &\textbf{133291}                        & \textbf{0.4132} & \textbf{2.1673} \\
        TopDown-B-ETS                  &\textbf{133271}                               & \textbf{0.4663} & \textbf{2.1810} \\
        Forecast reconciliation with minimum trace     &96261     & 0.1527          & 0.9861          \\
        Forecast reconciliation with weighted least squares  & 95922 & 0.1465          & 0.9961          \\
        \midrule
        \multicolumn{3}{c}{Panel B: errors of combination methods}                               \\
        \midrule
        Method                                               &             &             \\
        \midrule
        Simple average                  &84187                     & 0.1609          & 0.9468          \\
        Error-weighted average            &85293                   & 0.1816          & 1.1029          \\
        Stacking-lasso                      &79550                 & 0.0961          & 0.5421          \\
        Stacking-ridge                      &80361                 & 0.1051          & 0.5922          \\
        \bottomrule
      \end{tabular}
      }
\end{table}

\subsection{Reference forecasts}

In the second stage we estimate the reference forecasts from February 2023 to June 2023. Since the actual Chinese outbound tourism data during this period is available after the competition, we compare it with our estimations to evaluate the accuracy of our forecast framework. We further compare the effectiveness of search index and flight volume in predicting outbound tourism.

Table \ref{tab:table4} presents the Mean Absolute Scaled Error of the  reference forecasts and its two components, the reference forecasts estimated by flight numbers and the reference forecasts estimated by composite search index. For each destination, the lowest Mean Absolute Scaled Error among the three series is highlighted in bold to indicate the best-performing method. Since the direct flights from Mainland China to Chile, Mexico, Hawaii, and the Czech Republic have not resumed until June 2023 (or no direct flights have been offered), the reference forecasts of these four destinations are estimated solely based on the composite search index. The average and weighted average of Mean Absolute Scaled Error are also listed in this table, and weight is the average outbound tourism during February 2023 to June 2023.

On average, the reference forecasts based on composite search index yield the lowest average Mean Absolute Scaled Error of 1.6233, outperforming both the reference forecasts based on flight data (2.7868) and their mean value (1.7371). If weighted average value is set as the criterion, the mean value (2.0248) performs slightly better than search index (2.0779). In 16 destinations where predictions based on flight are available, the predictions based on search index perform best in 7 of them while the predictions based on flight perform best in only 3 of them. The average prediction performs best in 6 destinations. This suggests that the high frequency search behavior may be a more stable and reliable indicator in estimating tourism. Search data provides an immediate insight into people's preparation for their traveling, such as looking for destinations, checking visa policies, or ordering hotels, which is highly correlated with actual travel decisions. The flight data, however, does not maintain a very stable relation with outbound tourism in the highly uncertain situation in 2023. Since the travel restrictions have been lifted, the number of direct flights recover swiftly in a short period of time. However, the seat occupancy rate may not have recovered to an ordinary level, so we may overestimate the outbound tourism in this situation, as is the case of Maldives, Thailand, Turkey and Australia. Their Forecast error based on flight data is significantly higher than the error based on search data. So, we recommend using the search index when predicting the short-term trend of outbound tourism.

\begin{table}
  \centering
  \caption{Mean Absolute Scaled Error of reference forecasts and its components (from January 2023 to June 2023).}
  \label{tab:table4}
  \begin{tabular}{lccc}
    \toprule
    Destination      & Reference & Direct flight   & Composite search index \\
    \midrule
    Canada           & \textbf{0.4708}  & 0.8802          & 0.8085                 \\
    Chile            & 1.0963           & --              & 1.0963                 \\
    Mexico           & 0.6944           & --              & 0.6944                 \\
    Chinese Taipei   & 1.9542           & \textbf{1.3703} & 2.5381                 \\
    Hong Kong, China & 3.0323           & 3.8043          & \textbf{3.0323}        \\
    Japan            & 0.9430           & 0.9670          & \textbf{0.9190}        \\
    Korea (ROK)      & \textbf{0.2937}  & 1.1461          & 1.3760                 \\
    Macao, China     & \textbf{0.9888}  & 1.0328          & 1.1545                 \\
    Maldives         & 6.3710           & 11.2578         & \textbf{1.4841}        \\
    Cambodia         & 1.4194           & \textbf{1.3805} & 2.5733                 \\
    Indonesia        & \textbf{3.0946}  & 3.1808          & 4.4610                 \\
    Singapore        & \textbf{0.9567}  & 2.1413          & 1.1924                 \\
    New Zealand      & \textbf{1.5182}  & 1.7451          & 1.2914                 \\
    USA              & 1.0160           & \textbf{0.7334} & 1.3443                 \\
    Thailand         & 2.3467           & 5.5711          & \textbf{1.5206}        \\
    Turkey           & 1.4537           & 3.3289          & \textbf{0.8521}        \\
    Australia        & 2.1087           & 4.8322          & \textbf{1.2256}        \\
    Hawaii           & 1.6194           & --              & 1.6194                 \\
    Austria          & 0.9694           & 1.2176          & \textbf{0.8884}        \\
    Czech Republic   & 2.3953           & --              & 2.3953                 \\
    \midrule
    Average          & 1.7371           & 2.7868          & \textbf{1.6233}        \\
    Weighted Average   & \textbf{2.0248}           & 2.6610          & 2.0779        \\
    \bottomrule
  \end{tabular}
\end{table}

\subsection{Recovery curve forecasts}

In the third stage of RISE framework, three assumptions are made to estimate the recovery curve from August 2023 to July 2024 based on the base forecasts made in the first stage and the reference forecasts made in the second stage. We also set a recovery coefficient for each destination in order to capture the potential difference of recovery speed. According to the latest outbound tourism data disclosed after the competition (until January 2025), a retrospective analysis is made to evaluate the performance of forecasts.

Table \ref{tab:error_metrics} presents the Root Mean Squared Error, Mean Absolute Percentage Error, and Mean Absolute Scaled Error of the recovery forecasts for each destination. Overall, the RISE framework demonstrates satisfactory accuracy across the majority of destinations. The Mean Absolute Percentage Error values remain below 0.3 in 11 destinations. Destinations like Chinese Taipei, Hawaii, and Cambodia record relatively large errors, reflecting structural shifts in their post-pandemic tourism. These results highlight the heterogeneous nature of tourism recovery across markets and underscore the importance of incorporating more destination-specific factors into forecasting models.

\begin{table}[htbp]
    \centering
    \caption{Different forecast errors of recovery curve forecasts by destination (from August 2023 to July 2024).}
    \label{tab:error_metrics}
    \resizebox{\textwidth}{!}{
    \begin{tabular}{lccc}
        \toprule
        Destination      & Root Mean Squared Error     & Mean Absolute Percentage Error   & Mean Absolute Scaled Error   \\
        \midrule
        Canada           & 7195    & 0.2824 & 1.3871 \\
        Chile            & 878     & 0.3310 & 2.3311 \\
        Mexico           & 2726    & 0.1654 & 0.7668 \\
        Chinese Taipei   & 28703   & 0.9553 & 8.3389 \\
        Hong Kong, China & 1328443 & 0.4781 & 2.9475 \\
        Japan            & 48990   & 0.1214 & 0.8486 \\
        Korea (ROK)      & 39274   & 0.1296 & 1.3411 \\
        Macao, China     & 348843  & 0.1708 & 1.0657 \\
        Maldives         & 3808    & 0.1697 & 0.5389 \\
        Cambodia         & 39083   & 0.6768 & 5.2619 \\
        Indonesia        & 33477   & 0.3557 & 3.4175 \\
        Singapore        & 43290   & 0.1828 & 0.7896 \\
        New Zealand      & 5130    & 0.2388 & 0.8169 \\
        USA              & 27409   & 0.9146 & 0.6401 \\
        Thailand         & 103227  & 0.2464 & 1.4301 \\
        Turkey           & 10519   & 0.3357 & 2.7209 \\
        Australia        & 11712   & 0.1762 & 0.4904 \\
        Hawaii           & 2106    & 1.9698 & 5.0964 \\
        Austria          & 6987    & 0.2355 & 1.4804 \\
        Czech Republic   & 13025   & 0.5624 & 3.2130 \\
        \bottomrule
    \end{tabular}
    }
\end{table}

Table \ref{tab:table5} presents the Mean Absolute Scaled Error of the three components of recovery curve forecasts, the recovery curves based on linear, quadratic and logistic assumptions. For each destination, the lowest Mean Absolute Scaled Error among the three series is highlighted in bold to indicate the best-performing method. The simple average and weighted average of Mean Absolute Scaled Error is also listed in this table, and weight is the average outbound tourism during August 2023 to July 2024. Among three assumptions, the quadratic assumption achieves the best and most stable performance, its average Mean Absolute Scaled Error is 1.9906, significantly lower than other two methods. Among 20 destinations, linear method performs best in 8 destinations, however, it performs poorly in Indonesia and if we ignore that country, the performance of linear assumption and quadratic assumption should be similar. However, if we use weighted average value as the criteria, the logistic method performs best as its error in Hong Kong and Macao is fairly small.

\begin{table}
  \centering
  \caption{The Mean Absolute Scaled Error of recovery curve forecasts with respect to three curve functions (from August 2023 to July 2024).}
  \label{tab:table5}
  \begin{tabular}{lccc}
    \toprule
  Destination        & Linear          & Quadratic       & Logistic        \\
    \midrule
    Canada           & 1.5731          & \textbf{1.3975} & 1.4430          \\
    Chile            & \textbf{2.0116} & 2.1836          & 2.7980          \\
    Mexico           & \textbf{0.4868} & 0.5054          & 1.4691          \\
    Chinese Taipei   & 6.6705          & \textbf{5.2014} & 13.1448         \\
    Hong Kong, China & 3.7195          & 4.0265          & \textbf{1.1836} \\
    Japan            & 0.9979          & \textbf{0.9479} & 1.6394          \\
    Korea (ROK)      & \textbf{0.8470} & 1.6343          & 2.5268          \\
    Macao, China     & \textbf{0.9919} & 1.4723          & 1.0298          \\
    Maldives         & 1.2788          & \textbf{0.5194} & 0.9948          \\
    Cambodia         & 4.7207          & \textbf{3.9109} & 7.1542          \\
    Indonesia        & 6.4888          & 2.7449          & \textbf{1.5641} \\
    Singapore        & \textbf{0.5813} & 0.9811          & 0.8821          \\
    New Zealand      & \textbf{0.6338} & 0.9715          & 1.2857          \\
    USA              & 0.7420          & 0.8192          & \textbf{0.7203} \\
    Thailand         & 2.2996          & 1.0091          & \textbf{0.9816} \\
    Turkey           & \textbf{2.1860} & 2.4690          & 3.5079          \\
    Australia        & 0.6411          & \textbf{0.4525} & 0.6395          \\
    Hawaii           & 3.8336          & \textbf{3.3592} & 8.1253          \\
    Austria          & 2.4500          & 2.0315          & \textbf{1.7284} \\
    Czech Republic   & \textbf{3.1708} & 3.1744          & 3.4261          \\
    \midrule
    Average          & 2.3162          & \textbf{1.9906} & 2.8122          \\
    Weighted Average & 2.1719          & 2.1848          & \textbf{1.4904} \\
    \bottomrule
  \end{tabular}
\end{table}

Meanwhile, we have also compared the forecasting performance of the RISE framework with several benchmark models, including Na\"ive, Seasonal Na\"ive, Seasonal ARIMA and the exponential smoothing state space model in Table \ref{tab:benchmark}. The benchmark results in this paper are provided by the organizer of the tourism forecasting competition. As shown in Table \ref{tab:benchmark}, our RISE framework achieves the lowest overall Mean Absolute Scaled Error ($2.1464$), outperforming all benchmark methods by a substantial margin. The RISE framework achieves the best performance in $14$ out of $20$ destinations, except Chile, Chinese Taipei, Hong Kong, Indonesia, Hawaii and Czech. The Na\"ive and SARIMA model performs better than other benchmark models. In summary, the RISE framework integrates baseline forecasts with scenario-based recovery adjustments and cross-country heterogeneity through recovery coefficients, allowing it to adapt to different recovery patterns. This design enables RISE to balance flexibility and robustness, yielding the most stable and accurate results across different destinations.

\begin{table}
    \label{tab:benchmark}
    \centering
    \caption{The Mean Absolute Scaled Error of the RISE framework and benchmark methods (from August 2023 to July 2024).}
    \resizebox{\textwidth}{!}{
      \begin{tabular}{lccccc}
        \toprule
        Destination & Na\"ive & Seasonal Na\"ive & Seasonal ARIMA & ETS & RISE \\
        \midrule
        Canada                   & 1.9270 & 2.2533 & 1.5731 & 1.4396 & \textbf{1.3857} \\
        Chile                    & 2.0652 & 2.5264 & \textbf{1.6759} & 2.0652 & 2.1449 \\
        Mexico                   & 2.8898 & 3.2289 & 3.2209 & 2.4177 & \textbf{0.7096} \\
        Chinese Taipei           & 0.3527 & 0.4582 & \textbf{0.3265} & 0.3527 & 7.8455 \\
        Hong Kong, China         & 3.0908 & 4.9106 & \textbf{2.5622} & 7.1695 & 2.9527 \\
        Japan                    & 4.8940 & 5.0753 & 4.6105 & 4.8662 & \textbf{0.8395} \\
        Korea (ROK)              & 3.2140 & 3.4793 & 3.2424 & 3.1601 & \textbf{1.3647} \\
        Macao, China             & 3.0532 & 4.7326 & 3.0607 & 3.2233 & \textbf{1.0914} \\
        Maldives                 & 2.9887 & 4.8270 & 1.8294 & 4.2252 & \textbf{0.4965} \\
        Cambodia                 & \textbf{0.5247} & 1.9055 & 3.1346 & 0.8074 & 5.0514 \\
        Indonesia                & \textbf{1.4034} & 1.8393 & \textbf{1.4034} & \textbf{1.4034} & 3.1392 \\
        Singapore                & 3.7875 & 4.2067 & 2.9907 & 3.8484 & \textbf{0.7418} \\
        New Zealand              & 3.0157 & 3.4054 & 1.5726 & 3.1423 & \textbf{0.7526} \\
        USA                      & 2.7450 & 2.8772 & 2.7536 & 2.1046 & \textbf{0.6695} \\
        Thailand                 & 3.2220 & 4.3335 & 3.2220 & 3.2221 & \textbf{1.4000} \\
        Turkey                   & 4.5574 & 4.6531 & 4.6880 & 4.5561 & \textbf{2.5493} \\
        Australia                & 2.6870 & 5.4711 & 3.7795 & 4.5368 & \textbf{0.4837} \\
        Hawaii                   & 0.5458 & \textbf{0.2084} & 0.2788 & 0.5347 & 4.7944 \\
        Austria                  & 2.1259 & 2.1620 & 1.9962 & 1.8185 & \textbf{1.3802} \\
        Czech                    & 0.7876 & 0.9990 & \textbf{0.5670} & 1.4919 & 3.1358 \\
        \midrule
        Average                  & 2.4939 & 3.1776 & 2.4244 & 2.8193 & \textbf{2.1464} \\
        \bottomrule
      \end{tabular}
    }
\end{table}

\begin{figure}[h]
    \centering
    \includegraphics[width=\textwidth]{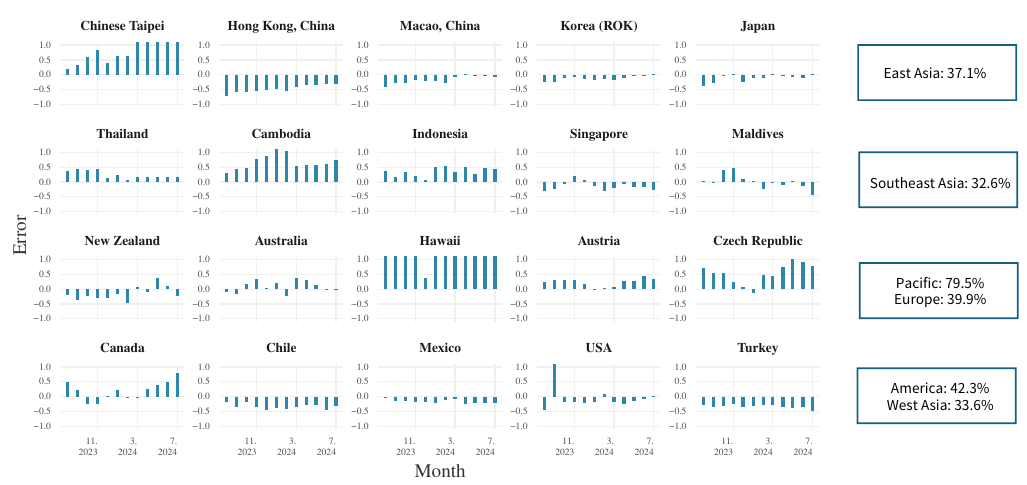}
    \caption{The percentage error of outbound tourism forecasts from August 2023 to July 2024.}
    \label{fig: mape-by-region}
\end{figure}

We utilize the \emph{percentage error} which is defined as $(F_{Prediction} - F_{Actual})/{F_{Actual}}$ to just quantify the deviation of a forecast from the actual value relative to the actual value itself. This measure indicates the direction and magnitude of forecast bias, where a positive value suggests overestimation and a negative value indicates underestimation. While intuitive and easy to interpret, this metric is sensitive to small actual values, which can lead to large or undefined errors, making it potentially unstable in practice. Figure \ref{fig: mape-by-region} demonstrates the \emph{percentage error} across different destinations from August 2023 to July 2024, revealing significant geographic heterogeneity in forecasting accuracy. Notably, the model tends to underestimate outbound tourism to nearby destinations such as Hong Kong, Macao, Japan, and Korea. This is likely due to the rapid recovery in short-haul travel, which outpaced the assumptions embedded in the recovery coefficients. In contrast, destinations in Europe and North America show overestimated forecasts, possibly attributable to lingering visa restrictions, incomplete flight restoration, and subdued long-haul travel sentiment. In Southeast Asia, the model performs well for countries like Singapore and the Maldives but overestimates demand to Thailand and Cambodia, potentially due to safety concerns. Forecast errors in Pacific and Latin American destinations are relatively low. Overall, Figure \ref{fig: mape-by-region} also suggests that the fixed recovery coefficients is effective in capturing the basic trend of tourism recovery but may not fully reflect rapid or asymmetric recovery trajectories.

\begin{figure}[h]
    \centering
    \includegraphics[width=1.0\textwidth]{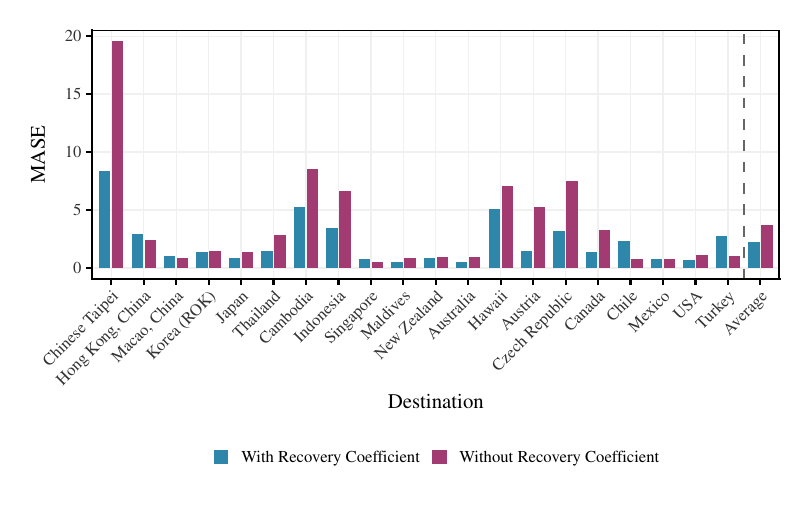}
    \caption{Mean Absolute Scaled Error with and without incorporating recovery coefficients.}
    \label{fig:fig3}
\end{figure}

In the third stage, a recovery coefficient is set for each destination considering the long-term influence of COVID-19. What if we remove the recovery coefficient in our forecast framework? Figure \ref{fig:fig3} shows the Mean Absolute Scaled Error of prediction with and without recovery coefficient. When removing the recovery coefficient, the average Mean Absolute Scaled Error increase to 3.6822 from 2.2461, and 15 out 20 countries/districts have higher Mean Absolute Scaled Error, indicating that the introduction of recovery coefficient significantly improves the forecast accuracy. However, in east Asian destinations, the recovery of tourism are far much faster than we have estimated previously, so the application of recovery coefficient are misleading in these destinations.

\subsection{Interval forecast}

Table~\ref{tab:interval} presents the evaluations of interval forecasts across 20 destinations using three performance measures: the Winkler score, the standard Winkler score, and the coverage probability as Equations (\ref{eq: winkler_total}) -- (\ref{eq: cp_total}).

The results show substantial variation in interval forecasts performance across destinations, reflecting heterogeneity in both the uncertainty and predictability of tourism demand recovery. For example, Canada, Mexico, and Korea (ROK) exhibit relatively low Winkler scores combined with coverage probabilities close to the nominal level, indicating that the intervals are not only sharp but also well-calibrated. In contrast, destinations such as Hong Kong, Cambodia, and Hawaii display much higher Winkler scores and considerably lower coverage probabilities, pointing to either excessively wide intervals or intervals that fail to adequately capture realized values. These results can largely be attributed to substantial errors in our model’s estimation of recovery rates for these destinations. The COVID-19 might incur  systematic shifts in these destinations, there by making it difficult for the interval forecasts to capture such changes.

\begin{table}[htbp]
  \centering
  \caption{Evaluation of interval forecasts using the Winkler score and coverage probability.}
  \label{tab:interval}
  \begin{tabular}{lccc}
    \toprule
    Destination     & Winkler Score & Standard Winkler Score & Coverage Probability \\
    \midrule
    Canada          & 26693         & 1.22                   & 58\%                 \\
    Chile           & 5023          & 1.98                   & 17\%                 \\
    Mexico          & 8589          & 0.52                   & 92\%                 \\
    Chinese Taipei   & 153664        & 5.26                   & 25\%                 \\
    Hong Kong China & 9881721       & 3.60                   & 0\%                  \\
    Japan           & 324513        & 0.73                   & 75\%                 \\
    Korea (ROK)     & 184305        & 0.56                   & 83\%                 \\
    Macao China     & 1744672       & 0.90                   & 67\%                 \\
    Maldives        & 20936         & 0.99                   & 75\%                 \\
    Cambodia        & 161546        & 2.80                   & 8\%                  \\
    Indonesia       & 94423         & 1.03                   & 58\%                 \\
    Singapore       & 179402        & 0.84                   & 67\%                 \\
    New Zealand     & 24355         & 1.21                   & 58\%                 \\
    USA             & 146048        & 1.28                   & 67\%                 \\
    Thailand        & 359414        & 0.74                   & 83\%                 \\
    Turkey          & 57454         & 1.89                   & 0\%                  \\
    Australia       & 43727         & 0.64                   & 92\%                 \\
    Hawaii          & 13199         & 10.96                  & 8\%                  \\
    Austria         & 20042         & 0.76                   & 92\%                 \\
    Czech           & 56897         & 2.81                   & 25\%                 \\
    \bottomrule
  \end{tabular}
\end{table}

\section{Conclusion and Discussions}
\label{sec: conclusion}

The COVID-19 pandemic has profoundly disrupted global travel, creating unprecedented levels of uncertainty and structural shifts that challenge traditional forecasting methods. In this paper, we develop a three-stage, recovery-informed forecasting framework designed to project the rebound of Chinese outbound tourism following the pandemic. This framework integrates multiple forecasting models trained on pre-pandemic data, real-time external indicators, and expert insights into recovery trajectories, offering a transparent and adaptable approach to forecasting in highly volatile environments. Results from the Tourism Forecasting Competition Round II validate the effectiveness of the proposed method, showing robust performance across most destinations and emphasizing the benefit of combining quantitative models with structured expert judgment.

This study advances the emerging literature on post-crisis tourism demand forecasting by illustrating how quantitative modeling can be effectively combined with expert judgment amidst profound uncertainty. The RISE framework synthesizes model-based projections with real-time external data and judgmentally assigned recovery coefficients, offering a systematic methodology for forecasting when traditional time series approaches are hindered by structural breaks. This approach balances statistical precision with contextual adaptability, aligning with the growing recognition that hybrid models outperform purely data-driven or solely subjective methods during periods of upheaval. Crucially, this integration is vital for recovery forecasts with limited historical data, enabling experts to supplement quantitative analysis with real-time insights.

From a practical perspective, our findings demonstrate significant heterogeneity in recovery trajectories across different destinations. Short-haul locations such as Hong Kong, Macao, Japan, and Korea experienced notably faster recoveries than anticipated, underscoring the critical role of geographical proximity and regional connectivity. Conversely, long-haul destinations including Europe and North America saw more gradual rebounds, likely due to persistent visa restrictions, elevated travel expenses, and limited flight capacities. These observations indicate that distance continues to be a key factor influencing the pace of tourism recovery in the post-COVID-19 landscape. Additionally, our forecasts offer valuable insights for travel agencies and policymakers. For travel agencies, recognizing varying recovery patterns across markets allows for more precise marketing efforts and efficient resource deployment. Agencies can temporarily scale back resources—such as marketing budgets and sales staff—in markets with slower recovery, while closely monitoring indicators like search volume trends and flight resumption data to identify optimal timing for resource reallocation. Conversely, agencies can direct increased promotional activities toward destinations demonstrating higher recovery potential by expanding marketing investments, enhancing partnerships with local suppliers and airlines, and creating customized tour packages to draw more visitors. For policymakers, our results aid in predicting tourism activity fluctuations, enabling proactive adjustments in fiscal support, infrastructure development, and labor policies to better accommodate ongoing recovery dynamics.

While our findings provide valuable insights, this study has several limitations that should be acknowledged. First, the recovery coefficients were heuristically assigned based on geographic distance, recovery situation, and policy factors, whereas economic determinants were not quantitatively modeled. According to classical demand theory, the income level of origin-country residents and the relative price of destinations are among the most important determinants of tourism demand. However, macroeconomic variables are often more volatile than political or geographical factors and may be unreliable to include without accurate estimation. Given the unstable and rapidly changing nature of tourism recovery after the pandemic, the relationship between macroeconomic indicators and tourism arrivals may also be complex and inconsistent to model. Future research could incorporate macroeconomic indicators such as real GDP growth, consumer confidence, and relative price indices to better estimate their influence on tourism recovery. Given that some macroeconomic variables, such as GDP or CPI, are typically provided at quarterly or annual frequency, their role is more suitable for informing medium-term recovery scenarios or calibration of recovery assumptions, rather than serving as a direct input in our framework.

Beyond tourism, we contend that the RISE framework offers a versatile approach for predicting sector recovery following sudden shocks--applicable to industries such as hospitality, aviation, energy, and retail, where limited data and structural volatility challenge conventional methods. By integrating quantitative modeling with expert judgment, RISE provides a flexible, transparent, and adaptable forecasting tool suited for high-uncertainty contexts.

This study also opens several promising avenues for future research. While the current framework focuses on point forecasts, extending the RISE framework to fully probabilistic forecasting would allow for richer uncertainty quantification. Techniques such as Bayesian model averaging, quantile regression forests, or simulation-based ensembles could be investigated to model the entire forecast distribution under high uncertainty. Future studies could also develop data-driven approaches to dynamically update the recovery coefficients as new observations or policy changes arise. Although we employed hierarchical forecasting, the current model treats destinations independently during the recovery adjustment phase. However, international tourism networks often exhibit spillover effects—recovery in one region may influence others through connecting flights, shared demand shocks, or coordinated policies. A natural progression would be to apply spatial-temporal or graph-based models to capture such dependencies. Building on the recovery heuristic concept, the RISE framework could further support scenario-based forecasts linked to specific policy or market assumptions. Coupling this with downstream optimization tasks (e.g., tourism promotion planning or flight scheduling) would transition the framework toward a decision-oriented forecasting paradigm.

\section*{Acknowledgment}

Feng Li (principal investigator), Taozhu Ruan, Kexin Shi, Yuchen Xue, and Yiming Zhong participated in the \emph{Tourism Forecasting Competition amid COVID-19 Round II} as the team \emph{KLLAB\_CN}. The method proposed in the competition ranked first place in point forecasting and third place in interval forecasting. This paper was drafted after the competition, with Feng Li and Taozhu Ruan leading the manuscript preparation and extended analysis. The authors thank the competition organizers for providing the platform and data. The authors are grateful to the editors and four anonymous reviewers for helpful comments that improved the contents of the paper. This research is supported by the National Natural Science Foundation of China (No. 72495123).

% The authors participated in the \emph{Tourism Forecasting Competition amid COVID-19 Round II} and the authors thank the organizing committee for providing the platform and data. The authors are grateful to the Editor, Associate Editor and four anonymous reviewers for helpful comments that improved the contents of the paper. This research is supported by the National Natural Science Foundation of China (No. 72495123).

\printbibliography

\end{document}